\documentclass[final,3p,times]{elsarticle}

\usepackage{lineno}

\journal{}

\usepackage{graphicx}
\usepackage{float}
\usepackage{subfig}
\usepackage{color}
\usepackage{amsmath}
\usepackage{amssymb}
\usepackage{xfrac}
\usepackage{array}
\usepackage{multicol}
\newcommand{\RN}[1]{%
  \textup{\lowercase\expandafter{\romannumeral#1}}%
}

\begin{document}

\begin{frontmatter}


\title{Adjoint-Based Sensitivity Analysis of Steady Char Burnout}
\author[label1]{Ahmed Hassan\corref{cor1}}
\ead{a.hassan@itv.rwth-aachen.de}
\cortext[cor1]{}
\author[label1,label2]{Taraneh Sayadi}
\author[label3]{Vincent Le Chenadec}
\author[label1]{Heinz Pitsch}
\author[label1]{Antonio Attili}

\address[label1]{Institute for Combustion Technology, Aachen University, Germany}
\address[label2]{Jean Le Rond d'Alembert Institute, Sorbonne University, France}
\address[label3]{Multiscale Modelling and Simulation Laboratory, Paris -- Est University, France}

\begin{abstract}

Simulations of pulverised coal combustion rely on various models, required in order to correctly approximate the flow, chemical reactions, and behavior of solid particles. These models, in turn, rely on multiple model parameters, which are determined through experiments or small-scale simulations and contain a certain level of uncertainty. The competing effects of transport, particle physics, and chemistry give rise to various scales and disparate dynamics, making it a very challenging problem to analyse. Therefore, the steady combustion process of a single solid particle is considered as a starting point for this study. As an added complication, the large number of parameters present in such simulations makes a purely forward approach to sensitivity analysis very expensive and almost infeasible. Therefore, the use of adjoint-based algorithms, to identify and quantify the underlying sensitivities and uncertainties, is proposed. This adjoint framework bears a great advantage in this case, where a large input space is analysed, since a single forward and backward sweep provides sensitivity information with respect to all parameters of interest. In order to investigate the applicability of such methods, both discrete and continuous adjoints are considered, and compared to the conventional approaches, such as finite differences, and forward sensitivity analysis. Various quantities of interest are considered, and sensitivities with respect to the relevant combustion parameters are reported for two different freestream compositions, describing air and oxy-atmospheres. This study serves as a benchmark for future research, where unsteady and finally turbulent cases will be considered.
\end{abstract}

\end{frontmatter}

\section{Introduction}

Most energy conversion systems operate by using liquid sprays or pulverized fuel in the combustion process, rendering the understanding of the interaction between droplets/particles essential in predicting the ignition, combustion behaviour, and even the formation of pollutants. Since these systems are composed of clusters of droplets/particles, it stands to reason that the understanding of the combustion process of isolated droplets/particles can lead to better predictions, forming the basis of many analytical, numerical, and modelling investigations~\cite{sirignano2010, Annamalai1993}. As far as coal particles are concerned, combustion starts with the devolatilization process, where partially oxidized smaller hydrocarbons and tars are released from the fuel particles leaving a char residue behind. Typically, devolatilization occurs on a much faster time scale than the char burnout. As a result, it is common practice to analyse the coal char combustion separately, which for coal particles with relatively small volatile matter is an acceptable assumption. For the purpose of this study, therefore, we will concentrate on char burnout. 

Simulations of such reactive systems are achieved by tackling the interaction of transport and chemistry. The governing equations are: (i) the equations associated with the description of the flow, involving velocities, pressure, and energy and (ii) the equations concerning the chemical evolution of the reactive species, involving the determination of the chemical reaction rates and corresponding distribution of the species mass fractions. In real flames, these two sets of equations are coupled. In the case of char burnout, in particular, the reactions on the surface of the particle (heterogenous reactions) should also be accounted for, adding to the overall complexity. However, due to the prohibitive costs of reactive flow computations, arising from the large range of scales and the equations governing the evolution of all the species, usually some level of modelling is employed. These models range from very low fidelity models such as single-film~\cite{Nusselt1924,Essenhigh1961}, double-film~\cite{Burke1931,Spalding1951}, and continuous-film models~\cite{Hecht2013}, to higher fidelity models used in direct numerical simulations~\cite{Lee1996,Higuera2008,Richter2013,Farazi2017}. A comprehensive study of char burnout models is given in Leiser~\cite{Leiser2011}.

Regardless of their fidelity, all existing models rely on many parameters in order to approximate the physical solution. These parameters are normally determined by experiments or small scale simulations and contain a certain level of uncertainty. Therefore, it is of great importance to determine the sensitivity of the output (final solution or quantity of interest) to the existing parameters in the model. The extracted sensitivities (uncertainties) can later be used to build models consistent with the experimental data, as is done in a study by Lavarone \textit{et al}~\cite{Lavarone2018}, where Bound-to-Bound Data Collaboration is used to derive a reduced char combustion model in both conventional and oxy-conditions.

In practice, most of the sensitivity information is extracted in zero-dimensional or one-dimensional settings, focusing solely on the kinetics, and therefore ignoring the effect of the underlying multi-dimensional flow~\cite{Rabitz1983,Saltelli2005,Cai 2018}. This is usually due to the fact that extracting the Jacobian (gradient) of the full system, including the coupled hydrodynamic and reactive equations, using conventional approaches, becomes very expensive and in many real-scale applications infeasible.

Common methods in extracting the gradient information are analytical or use finite differences. As far as the analytical determination of the gradient is concerned, the subtleties associated with the numerical approach, such as the meshing process, discretization, and the complexity of the governing equations, make extracting the analytical gradient expression far from straightforward. In the case of finite differences, on the other hand, the numerous parameters involved in modelling make this approach prohibitively expensive for large problems and also overwhelmed by numerical noise. Adjoint-based methodologies allow the determination of the gradient at a cost comparable to a single function evaluation, regardless of the number of parameters, and are, therefore, a suitable alternative~\cite{Giles2000}. The backward solution (the solution of the adjoint equation) provides gradient or sensitivity information about an optimization problem. This gradient is computed in the form of algebraic expressions based on the problem's Lagrange multipliers or adjoint variables, which, in turn is used in standard optimization algorithms that rely on Jacobian information (such as the conjugate-gradient family). The use of adjoint methods for design and optimization has been an active area of research, which started with the pioneering work of Pironneau~\cite{Pironneau1974}, with applications in fluid mechanics, and later in aeronautical shape optimization by Jameson and co-workers~\cite{Jameson1988,Jameson1998,Reuther1999}. Ever since these ground-breaking studies, adjoint-based methods have been widely used in fluid mechanics, especially in the field of acoustics and thermo-acoustics~\cite{Juniper2010,Lemke2013}. These areas provide a suitable application for adjoint-based methods, which are inherently linear, since they too are dominated by linear dynamics. Recently, nonlinear problems have also been tackled, within the context of optimal control of separation on a realistic high-lift airfoil or wing, enhancement of mixing efficiency, and minimal turbulence seeds, to name a few~\cite{Schmidt2013,Rabin2014,Foures2014}. In the case of reactive flows, adjoint-based methods have also been employed for adaptive mesh refinement in steady-state Reynolds Averaged Navier-Stokes (RANS) simulations~\cite{Duraisamy2012}. As far as detailed simulations are concerned, Braman \textit{et al.}~\cite{Braman2015} have investigated the applicability of the adjoint-based techniques in one- and two-dimensional laminar flow configurations, where the hydrodynamics are decoupled from the species transport equations. In a recent work, Lemke \textit{et al.}~\cite{Lemke2019} have also explored the applicability of such techniques, in the zero-dimensional case of a homogeneous constant volume reactor, to study the sensitivity of the ignition delay time with respect to the parameters appearing in the combustion model. Adjoint-based methods have also been applied in the linear limit in various studies~\cite{Sashittal2016,Skene2019,Capecelatro2017}, using simplified combustion models. Although these recent efforts show great promise in integrating adjoint-based methodologies in combustion technology for the purpose of sensitivity analysis and control, there still remain many open questions, regarding (i) the effect of nonlinearities and unsteadiness (commonplace in realistic reactive configurations), (ii) the coupling of the hydrodynamics to the chemistry, and (iii) non-conventional boundary conditions (such as in the presence of evaporation or surface reactions), on the extraction and performance of the adjoint equations. 

Therefore, the two main objectives of this study are (i) to extract a general formulation for the adjoint problem, applied to fully coupled detailed simulations of steady char particle burnout, in the nonlinear limit, for the purpose of sensitivity analysis with respect to the existing model variables, using a flexible approach, which is easily extendable to higher dimensions; and (ii) to compare continuous and discrete adjoint formulations in order to investigate the most advantageous strategy for gradient extraction, as far as reactive flows with non-trivial boundary conditions, nonlinear dynamics, and fully-coupled hydrodynamic/reactive equations, are concerned. These goals are pursued with a model proposed by Matalon~\cite{matalon1980}, allowing the extraction of the exact solution for the primal and adjoint problems, and facilitating the assessment of continuous versus discrete adjoints. This work, therefore, is to be considered as the first step of developing an adjoint-based framework, that will be extended to cases with unsteady dynamics, including detailed kinetic mechanism, and a subject of the future research.

This manuscript is organized as follows. Section~\ref{sec:GE} describes the equations governing the homogeneous reactions in the gas-phase and heterogeneous reactions appearing on the surface of the particle, that together form the primal (forward) problem. The adjoint-based framework is then presented in section~\ref{sec:Adj}, as well as the formulation for the extracted sensitivities. The numerical framework for the solution of both primal and adjoint problems are validated in section~\ref{sec:val} and the results are presented and discussed in section~\ref{sec:results}. Conclusions of the work are discussed in section~\ref{sec:con}, and the formulation details required to reproduce the presented results are finally given in~\ref{app:adj} and~\ref{app:bc}.

\section{Governing equations} \label{sec:GE}

Assuming a coal particle immersed in air, combustion is described by the following heterogeneous reactions on the surface of the particle (referred to as \emph{direct} and \emph{indirect} oxidation, respectively),
\begin{eqnarray}
\mathrm{C} \left [ s \right ] + \frac{1}{2} \mathrm{O}_2 \rightarrow \mathrm{CO}, \label{R1} \\ 
\mathrm{C} \left [ s \right ] + \mathrm{CO}_2  \rightarrow 2 \mathrm{CO}. \label{R2}
\end{eqnarray}
The following one-step mechanism is then considered for the oxidation of carbon monoxide in the gas phase,
\begin{equation}
\textrm{CO} + \dfrac{1}{2}\textrm{O}_2  \rightarrow \textrm{CO}_2. \label{R3} 
\end{equation} 
The evolution of the gaseous composition is described by the species mass fraction equations,
\begin{equation}
\nabla \cdot \left [ \rho \mathrm{Y_i} \left ( \mathbf{v} + \mathbf{V}_i \right ) \right ] = \omega_i,
\label{eq:species_org}
\end{equation}
where $i = 1, \cdots, 4$, refer to fuel ($\mathrm{CO}$), oxidizer ($\mathrm{O}_2$), product ($\mathrm{CO}_2$), and inert gas ($\mathrm{N}_2$), respectively. $\rho$ denotes the mixture density, $\mathbf{v}$ the mass-averaged velocity, and $\mathbf{V}_i$ the diffusion velocity for species $i$. The term on the right hand side of the above equation is, $\omega_i = \mathcal{W}_i (\nu_i'' - \nu'_i) \omega$, where $\nu''_i$ and $\nu'_i$ are the stoichiometric coefficients of the product and reactants in the homogeneous reaction, respectively, while $\mathcal{W}_i$ denotes the molecular weight of species $i$. The homogeneous reaction rate obeys an Arrhenius law~\cite{Howard1973},
\begin{equation}
    \begin{aligned}
	\omega & = k \left [ \mathrm{CO} \right ] \left [ \mathrm{O}_2 \right ]^{\sfrac{1}{2}} \left [ \mathrm{H}_2 \mathrm{O} \right ]^{\sfrac{1}{2}} \exp \left ( - \frac{E}{R T} \right ), \\
	& = \frac{k \mathrm{Y}_{\mathrm{H}_2 \mathrm{O}}^{\sfrac{1}{2}}}{\mathcal{W}_{\mathrm{CO}} \mathcal{W}_{\mathrm{O}_2}^{\sfrac{1}{2}} \mathcal{W}_{\mathrm{H}_2 \mathrm{O}}^{\sfrac{1}{2}}} \rho^2\mathrm{\mathrm{Y_{\mathrm{CO}} Y_{\mathrm{O}_2}}}^{\sfrac{1}{2}} \exp \left ( - \frac{E}{R T} \right ).
	\end{aligned}
	\label{eq:ReactionRate}
\end{equation}
Using Fick's law, the diffusion velocity for species $i$ can be closed as,
\begin{equation}
\mathrm{Y_i} \mathbf{V}_i = - D \nabla \mathrm{Y_i},
\label{eg:Fick}
\end{equation}
where the diffusion coefficient is assumed identical for all species. Substituting the previous equation in Eq.~\ref{eq:species_org} leads to the following transport equation for the species,
\begin{equation}
\nabla \cdot \left ( \rho \mathbf{v} \mathrm{Y_i} - \rho D \nabla \mathrm{Y_i} \right ) = \omega_i.
\end{equation}  
Similarly, assuming identical and constant specific heats, $C_p$, for all species, the energy equation yields the following equation for the temperature in the gas phase,
\begin{equation}
\nabla \cdot \left ( \rho C_p \mathbf{v} T - \lambda \nabla T \right ) = - \sum_{i = 1}^4 h^0_i \omega_i,
\end{equation}
where $h_i^0$ is the enthalpy of formation of species $i$,  and $\lambda$ is the coefficient of heat conduction. The low Mach number approximation is used. Finally, the continuity equation holds,
\begin{equation}
\nabla \cdot \left ( \rho \mathbf{v} \right ) = 0.
\end{equation}

For the purpose of this study, $\rho D$ is assumed constant and the Lewis number, $\mathrm{Le} = \lambda /\rho D C_p$ is set to unity. The variables are made dimensionless by introducing the following characteristic values, as proposed by Matalon \cite{matalon1980}: (i) the particle radius $a$ for length, (ii) $Q/C_p$ for temperature, where $Q$ is the heat of the homogenous reaction per unit mass of $\mathrm{CO}_2$, namely,
\begin{equation}
Q = - \frac{\sum_{i = 1}^4 h_i^0 \mathcal{W}_i \left ( \nu_i'' - \nu_i' \right )}{\mathcal{W}_{\mathrm{CO}_2} \left ( \nu''_{\mathrm{CO}_2} - \nu'_{\mathrm{CO}_2} \right )},
\end{equation}
(iii) $\rho_0 = p_0 C_p / Q \tilde{R}$, where $\tilde{R}$ is the specific gas constant based on an average molecular weight, $p_0$ is the uniform pressure, and (iv) $\lambda/\rho_0 C_p a$, for velocity. 
This non-dimensionalisation yields the following expression for the Damk\"ohler number of the homogeneous reaction:
\begin{equation}
 \mathrm{Da}^g= \frac{k \left ( \nu_{\mathrm{CO}_2}'' - \nu_{\mathrm{CO}_2}' \right ) C_p^3}{\lambda Q^2 \overline{R}^2} \frac{\mathcal{W}_{\mathrm{CO}_2}}{\mathcal{W}_{\mathrm{CO}} \mathcal{W}_{\mathrm{O}_2}^{\sfrac{1}{2}} \mathcal{W}_{\mathrm{H}_2 \mathrm{O}}^{\sfrac{1}{2}}} a^2 p_0^2 \mathrm{Y}_{\mathrm{H}_2 \mathrm{O}}^{\sfrac{1}{2}}.
\end{equation}
In non-dimensional form, the equation of state (ideal gas) reads $\rho T = 1$, and the (dimensionalized) governing equations are
\begin{equation}
\begin{aligned}
	\forall i \in \left \llbracket 1, 4 \right \rrbracket, \mathcal{L} \left ( \mathrm{Y_i} \right ) & = \Omega \alpha_i, \\
	\mathcal{L} \left ( \mathrm{T} \right ) & = \Omega \alpha_5,
\end{aligned}
\end{equation}
where, $\left ( \alpha_i \right )_{i \in \left \llbracket 1, 5 \right \rrbracket} = \left ( \alpha - 1, - \alpha, 1, 0, 1 \right )$, and $\alpha = \mathcal{W}_{\mathrm{O}_2} / 2 \mathcal{W}_{\mathrm{CO}_2}$, with the transport operator defined as
\begin{equation}
\mathcal{L} \left ( \varphi \right ) = \nabla \cdot \left ( \rho \mathbf{v} \varphi - \nabla \varphi \right ),
\end{equation}
and
\begin{equation}
\Omega = \mathrm{Da}^g \frac{\mathrm{Y_\mathrm{CO} Y_{\mathrm{O}_2}}^{\sfrac{1}{2}}}{T^2} \exp \left ( - \frac{\theta}{T} \right ).
\label{eq: final}
\end{equation}
If axisymmetry is assumed, the governing equations reduce to a set of coupled second order ordinary differential equations (ODE). In the case of a single spherical particle in particular, the continuity equation (in spherical coordinates) yields
\begin{equation}
r^2\rho v_r =\dot{ M} = \textrm{constant},
\end{equation}
where $r$ is the radial coordinate, $v_r$ the radial velocity component, and $\dot{M}$ denotes the ``burning rate''. The linear operator $\mathcal{L}$ then reduces to:
\begin{equation}
\mathcal{L}_{\mathrm{spherical}} \left ( \varphi \right ) = \frac{\dot{M}}{r^2} \frac{\mathrm{d} \varphi}{\mathrm{d} r} - \frac{1}{r^2} \frac{\mathrm{d}}{\mathrm{d} r} \left ( r^2 \frac{\mathrm{d} \varphi}{\mathrm{d} r} \right ).
\end{equation}
\subsection{Boundary Conditions} \label{ss:BC}
Away from the particle, composition ($\mathrm{Y}_{\mathrm{CO}}^{\infty}$, $\mathrm{Y}_{\mathrm{O}_2}^{\infty}$, $\mathrm{Y}_{\mathrm{CO}_2}^{\infty}$ and $\mathrm{Y}_{\mathrm{N}_2}^{\infty}$) and temperature ($T^\infty$) are set.
The particle is assumed isothermal ($\mathrm{T} = \mathrm{T_s}$ for $r < 1$), and the temperature profile is also assumed continuous, which yields the following boundary condition:
\begin{equation}
	\mathrm{T} \left ( r = 1 \right ) = \mathrm{T_s}.
\end{equation}

The remaining conditions at the particle surface are extracted from the conservation statements for each species $i$, which match the differences between the net fluxes of each species to the rates of consumption/production per unit area due to the heterogeneous reaction rates. Due to large solid to gas density ratio, the quasi-steady assumption holds, resulting in the following simplified mass balance for the $i^{th}$ species
\begin{equation}
	\rho \mathbf{v} \cdot \mathbf{n} \mathrm{Y_i}\mathrm{Y_{i}^{s}} - \rho D \left . \nabla \mathrm{Y_i} \right \vert^s \cdot \mathbf{n} = \omega'_i,
\end{equation}
where $\mathbf{n}$ denotes the outward-pointing normal to the particle surface, and $\omega'_i$ their respective reaction rate per unit area, due to the heterogeneous reactions. The boundary conditions are non-dimensionalized similarly to the governing equations variables.
The non-dimensionalized boundary conditions in axisymmetric form are then given as,
\begin{equation}
\dot{M} \mathrm{Y_{i}^{s}} - \left . \frac{\mathrm{d} \mathrm{Y_i}}{\mathrm{d} r} \right \vert^s = \omega_i'.
\end{equation}

Assuming both heterogeneous reactions are first order~\cite{matalon1980}, the following two heterogeneous Damk\"ohler numbers are defined,
\begin{equation}
\label{eq:Das}
\forall i \in \left \llbracket 1, 2 \right \rrbracket, \mathrm{Da}^s_i = \frac{C_p^2}{\lambda Q \overline{R}} a p_0 \frac{k_i \left ( T^s \right )}{T^s},
\end{equation}
for the direct ($i = 1$) and indirect ($i = 2$) oxidation reactions. This yields the conditions:
\begin{equation}
\dot{M} \mathrm{Y_{i}^{s}} - \left . \frac{\mathrm{d} \mathrm{Y_i}}{\mathrm{d} r} \right \vert^s =  \mathrm{Da}^s_1 \beta_i Y_{\mathrm{O}_2}^s +  \mathrm{Da}^s_2 \gamma_i Y_{\mathrm{CO}_2}^s,
\end{equation}
where $\left ( \beta_i \right )_{i \in \left \llbracket 1, 4 \right \rrbracket} = \left ( \left ( 1 - \alpha \right ) / \alpha, - 1, 0, 0 \right )$ and $\left ( \gamma_i \right )_{i \in \left \llbracket 1, 4 \right \rrbracket} = \left ( 2 \left ( 1 - \alpha \right ), 0, - 1, 0 \right )$.

\section{Two-point boundary value problem and adjoint-based gradient computation}
\label{sec:Adj}

In this section, the equations governing the char particle combustion, previously described in detail, are presented in the form of a more general two-point boundary value problem, and the corresponding continuous and discrete adjoint formulations are then presented.  

\subsection{Two-point boundary value problem}

The problem delineated above forms a two-point boundary value problem, the sensitivities of which are to be computed using the adjoint method.
To derive the adjoint equations, the methodology proposed by Serban \& Petzold~\cite{Serban2003} is used.

Let the governing equations be recast in first order form: $\mathbf{u} \left ( r \right ) \in \mathbb{R} ^ m$ ($m \in \mathbb{N}$) denotes the dependent variable, which includes species mass fraction and temperature derivatives as follows:
$$
\mathbf{u} = \left (\mathrm{ Y}_\mathrm{CO}, \mathrm{Y}_{\mathrm{O}_2}, \mathrm{Y}_{\mathrm{CO}_2}, \mathrm{Y}_{\mathrm{N}_2}, T, \dot{M}, \frac{\mathrm{d} \mathrm{Y}_\mathrm{CO}}{\mathrm{d} r}, \frac{\mathrm{d} \mathrm{Y}_{\mathrm{O}_2}}{\mathrm{d} r}, \frac{\mathrm{d} \mathrm{Y}_{\mathrm{CO}_2}}{\mathrm{d} r}, \frac{\mathrm{d} \mathrm{Y}_{\mathrm{N}_2}}{\mathrm{d} r}, \frac{\mathrm{d} T}{\mathrm{d} r} \right ).
$$

The solution of the problem delineated in the previous section is then obtained by solving a two-point boundary value problem of the form:
$$
\frac{\mathrm{d} \mathbf{u}}{\mathrm{d} r} = \mathbf{F} \left ( r, \mathbf{u}, \mathbf{p} \right ),
$$
to be solved from $r = a = 1$ (the particle surface) to $r = b = \infty$ subjected to the boundary conditions
$$
\mathbf{h} \left ( \mathbf{u} \left ( a \right ), \mathbf{u} \left ( b \right ), \mathbf{p} \right ) = 0.
$$
$\mathbf{p}$ denotes the problem parameters, that include the mixture composition away from the particle surface or the Damk\"{o}hler numbers in the case of interest here.

For the problem to be well-posed, the boundary condition operator $\mathbf{h}$ must provide $m$ independent boundary conditions. This is satisfied as long as the Jacobian,
\begin{equation}
\mathbf{h}_\mathbf{u} = \left [  \mathbf{h}_{\mathbf{u} \left ( a \right )} \vert \mathbf{h}_{\mathbf{u} \left ( b \right )} \right ] = \left [ \mathbf{A} \vert \mathbf{B} \right ],
\label{eqn:bcsensitivity}
\end{equation}
has full rank ($\mathbf{h}_{\mathbf{u} \left ( a \right )} = \mathbf{A}$ denotes the Jacobian of $\mathbf{h}$ w.r.t. $\mathbf{u} \left ( a \right )$, likewise for $\mathbf{h}_{\mathbf{u} \left ( b \right )} = \mathbf{B}$ and $\mathbf{u} \left ( b \right )$). The matrices $\mathbf{A}$ and $\mathbf{B}$ that result from the problem of interest are included in \ref{app:bc}.

In its first-order form, the two-point boundary value problem may be solved numerically using well-documented techniques, such as collocation or shooting methods. The latter was used in this work: it consists in solving the original problem as an initial value problem using readily available ODE integrators. The integration is performed outward from the particle surface but, since initial data such as mixture composition are missing, it is estimated and iteratively updated until the far-field conditions are matched. The iterative solver employed in this work is the Levenberg-Marquardt algorithm, which proved able to handle all the cases presented in the results section.

\subsection{Continuous adjoint formulation}

This section describes the derivation of the continuous adjoint equations, solved in the differentiate-then-discretise approach to adjoint-based gradient computation. The methodology followed here was proposed by Serban \& Petzold: it focuses on gradient of two types of functionals:
\begin{itemize}
\item Local (or pointwise) quantities, of the form $g \left ( s, \mathbf{u}, \mathbf{p} \right )$, where $s \in \left [ a, b \right ]$. This approach will be applied to the computation of the gradient of the burning rate at the particle surface with respect to model parameters ($s = a$).
\item Integral quantities, of the form:
$$
G^s \left ( \mathbf{p} \right ) = \int_a^s g \left ( r, \mathbf{u}, \mathbf{p} \right ) \mathrm{d} r.
$$
One application is the computation of the gradient of the integrated heat release ($s = b$).
\end{itemize}

The two problems are related by the Leibniz integral rule:
\begin{equation}
\frac{\mathrm{d}}{\mathrm{d} s} \frac{\mathrm{d} G ^ s}{\mathrm{d} \mathbf{p}} = \left . \frac{\mathrm{d} g}{\mathrm{d} \mathbf{p}} \right \vert _ s,
\label{eq:leibniz}
\end{equation}
or stated otherwise, the gradient of $g$ matches the sensitivity of the gradient of $G ^ s$ with respect to $s$.
As a result, both gradients can be computed by the adjoint method as follows:

\begin{enumerate}
\item Solution of the primal two-point boundary value problem, by means of shooting or collocation methods, as highlighted above.
\item Solution of the adjoint of the primal problem. This is a linear multi-point boundary value problem, which, in the limits $s \to a$ and $s \to b$ degenerates into a linear two-point boundary value problem.
\end{enumerate}

As shown by Serban \& Perzold, these two steps are sufficient to compute
\begin{equation}
\frac{\mathrm{d} G ^ s}{\mathrm{d} \mathbf{p}} = \int_a^s \left ( g_\mathbf{u} \mathbf{u}_\mathbf{p} + g_\mathbf{p} \right ) \mathrm{d} r.
\label{eq:integralgrad}
\end{equation}
In the event where the gradient of the local quantity $g$ evaluated at $s$ is of interest, the second step is replaced by:

\begin{enumerate}
\setcounter{enumi}{1}
\item Solution of the forward sensitivity of the adjoint problem with respect to $s$. This step gives rise to another linear multi-point boundary value problem that too degenerates into a linear two-point boundary value problem in the limits $s \to a$ and $s \to b$.
\end{enumerate}

For the sake of completeness, \ref{app:adj} delineates the procedure in depth.

\subsection{Discrete adjoint formulation}

Instead of deriving the continuous adjoint equations, an alternative approach, referred to as the discretise-then-differentiate approach, consists in extracting the adjoint equations directly from the discretised primal problem, where algorithmic differentiation is used to linearize and conjugate transpose the equations to ultimately yield the discrete adjoint formulation.

Since the discrete adjoint equations are extracted from the primal problem after the discretisation is performed, the adjoint and the primal grids are identical. This is one of the main differences between the continuous and the discrete approaches.

\section{Validation}
\label{sec:val}
In this section, the primal problem is first presented and validated in the frozen limit against analytical solutions. Once validated, sensitivities are extracted using the approaches to gradient computation discussed previously for a selected configuration. The advantages of 0.
each approach are highlighted in order to select the most suitable strategy. The following parameters are taken from the work of Matalon~\cite{matalon1980}: the homogeneous reaction frequency is set to $k=1.3 \times10^{14}$  [\textit{mole/ml.sec}], the pressure range to $1 \leq p_0 \leq 10$ [\textit{atm}], and the water mass fraction to $\mathrm{Y}_\mathrm{H_2O} \approx 10^{-3}$. The particle size, $a$, ranges between $10\; [\mu m]$ and $100 \; [\mu m]$. Using these values, the expression for $\mathrm{Da}^g$ becomes, $\mathrm{Da}^g =4.2 {\mathrm{Y}_\mathrm{H_2O}}^{\frac{1}{2}} a^2 p_0^2$.
 
\subsection{Primal problem}
\begin{figure}
\centering
  \subfloat[Varying surface Damk\"ohler numbers $ \mathrm{Da}^s_2$]{\includegraphics[width=0.45\textwidth]{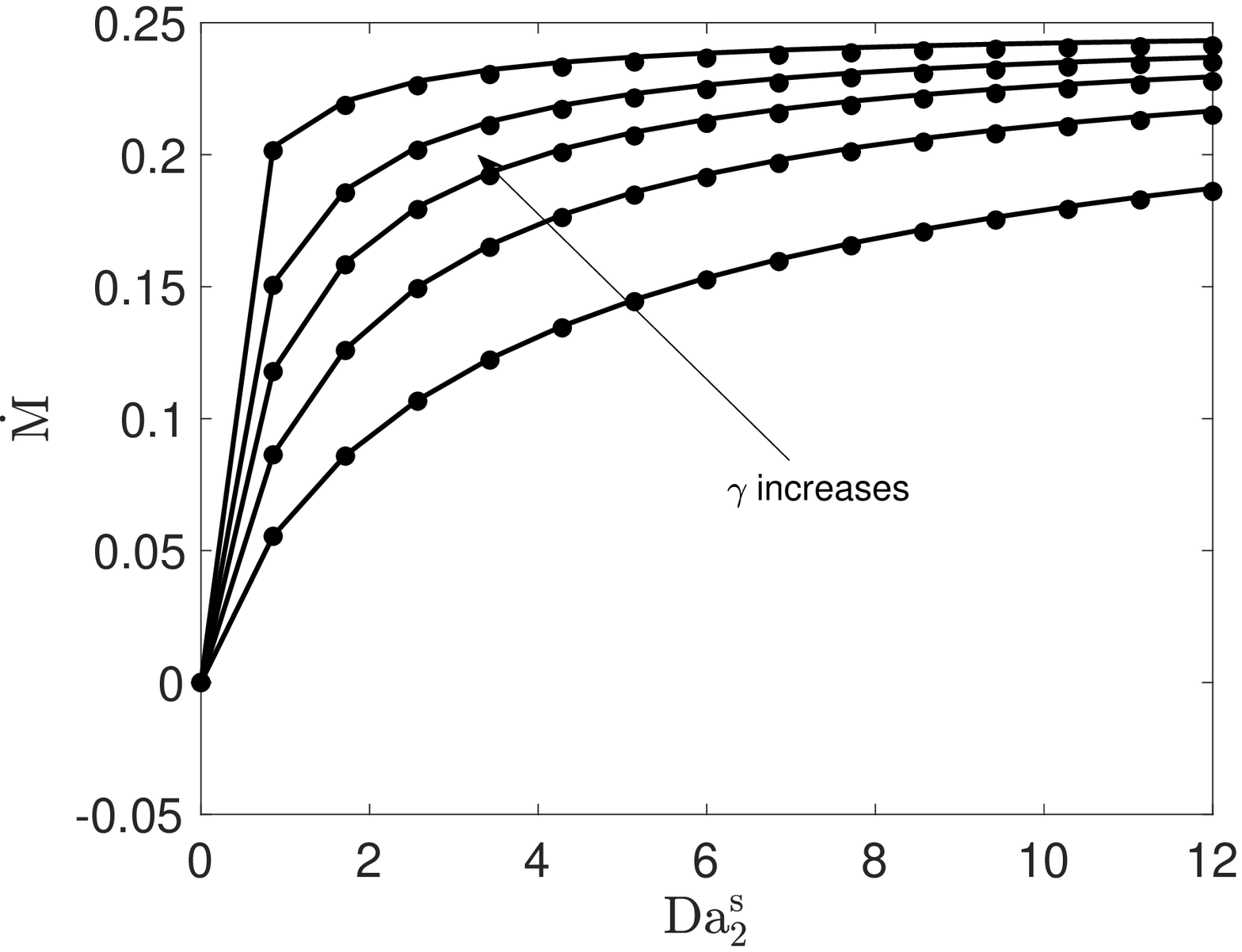}}
  \subfloat[Varying surface temperature $\mathrm{T_s}$]{\includegraphics[width=0.45\textwidth]{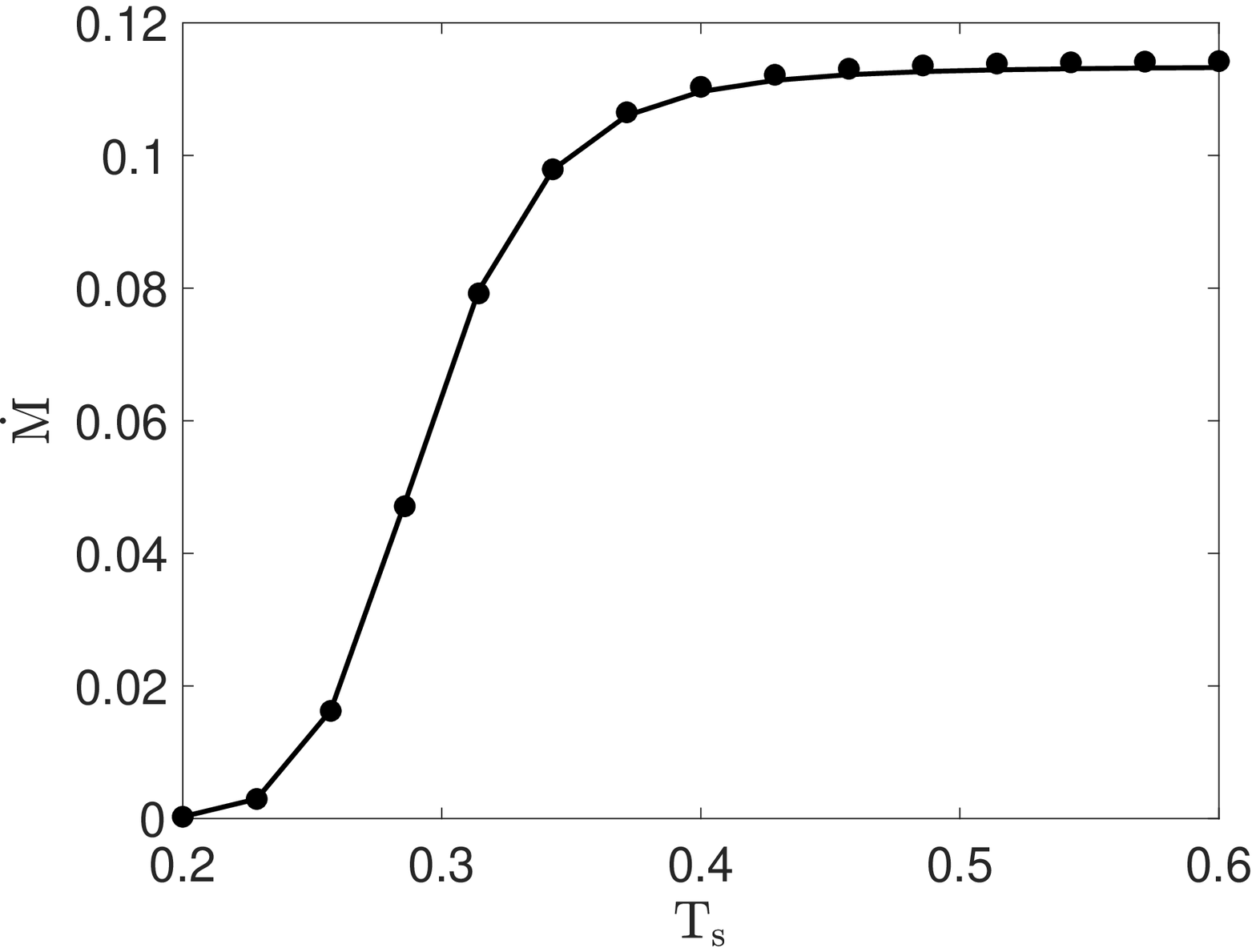}}
\caption{Comparison of the burning rate, predicted by the simulations, to the theoretical solution, in the frozen limit. ---, analytical solution; $\bullet$, numerical solution.}
\label{fig:figure1}
 \end{figure}

In this section, the solution of the primal problem is compared to the analytical solution in the frozen limit, where $\mathrm{Da}^g = 0$. The values of mass fractions and temperature in the far-field are set to $\mathrm{Y}_\mathrm{CO}^{\infty}= 0.05$, $\mathrm{Y}_{\mathrm{CO}_2}^{\infty}= 0.2$, $\mathrm{Y}_{\mathrm{O}_2}^{\infty}= 0.3$, $\mathrm{Y}_{\mathrm{N}_2}^{\infty}= 0.45$, ${T}^{\infty}= 0.056$ and $\gamma= \mathrm{Da}^s_2 / \mathrm{Da}^s_1 = 0.1$, as suggested by Matalon~\cite{matalon1980,matalon1981}. The far field temperature is set to $360\; [K]$, similar to the case studied by Matalon~\cite{matalon1980}. The dependency of the burning rate on $ \mathrm{Da}^s_1$ in the frozen limit ($\mathrm{Da}^g=0$), is shown in figure~\ref{fig:figure1}(a), for $0.1 \leq \gamma \leq 5$, and is compared to the analytical value. A comparison of approximated value for the burning rate, computed for different domain sizes (not shown here), showed that $L = 120a$ is the smallest length beyond which the solution becomes approximately independent of the domain size. However, figure~\ref{fig:figure1}(a) shows that increasing the Damk\"ohler number increases the error in the approximated solution. Changing the Damk\"ohler number changes the diffusion property of the solution, and augments the effect of the far field boundary conditions. Increasing the domain size for higher Damk\"ohler numbers reduces the resulting error. However, for the purpose of this study, since the resulting errors are in the acceptable range, the domain size is kept identical for all cases, and henceforth all the computations are performed on domain sizes $L \ge 120a$. As predicted by Matalon~\cite{matalon1980}, for small $\mathrm{Da}^s_1$, the surface reactions are very weak and the burning rate tends towards zero. The value of the burning rate, $\dot{M}$, on the other hand, increases with $\mathrm{Da}^s_1$ and moves asymptotically to $\dot{M}^\infty = 0.2432$. Figure~\ref{fig:figure1}(b) shows the dependence of the burning rate on the particle surface temperature $\mathrm{T_s}$. The predictions of the simulations are compared to the explicit expression derived by Makino~\cite{Makino1992}, where $T^\infty = 0.25$, $\mathrm{Y}_{\textrm{O}_2}^\infty = 0.1602$, $\mathrm{Y}_{\textrm{CO}_2}^\infty = 0$ and $\mathrm{Y}_{\textrm{N}_2}^\infty = 0.7398$. This expression is valid for burning rates sufficiently smaller than unity, an assumption which holds for the set of parameters chosen for this comparison. 

\subsection{Extracted sensitivities}

In this section, both the continuous and discrete adjoint methodologies are validated against the extracted sensitivities from the analytical solution of the primal problem, and the most suitable strategy, for applications of interest to this study, is identified. The accuracy of the adjoint methods are also compared to conventional methods such as finite difference and forward sensitivity.

\begin{figure}
\centering
 {\includegraphics[width=0.5\textwidth]{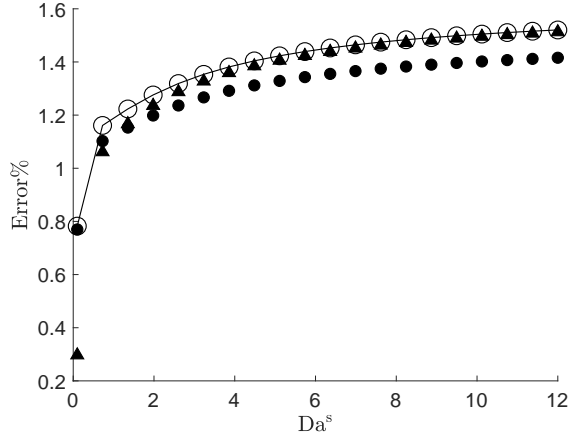}}
\caption{Error comparison of burning rate sensitivity with respect to $\mathrm{{Da}
_1^s}$.
$\bullet$, discrete adjoint; ---, continuous adjoint; $\blacktriangle$, finite difference; $\circ$, forward sensitivity. Error computed with respect to the analytical solution of the adjoint problem.}
    \label{fig:figure2}
 \end{figure}

Figure~\ref{fig:figure2} compares the error in the estimated sensitivities. The values of mass fractions and temperature in the far stream are identical to the ones previously used. The results are reported for $\gamma=0.1$. A high value of $\gamma$ results in higher flame temperatures, leading to larger gradients in the species and temperature fields, and resulting in more clear comparisons. While all methods result in similar values for the extracted sensitivity, the discrete adjoint methodology proves to be the most accurate. This accuracy increases with the surface Damk\"ohler number. Increasing the surface  Damk\"ohler number results in a solution closer to the asymptotic limit for the primal problem, which can be seen in figure~\ref{fig:figure1}(a). In this region, the accuracy of the forward solution also decreases, showing higher errors compared to the exact values. As a result, the increase in the estimated errors of the gradient can be attributed to the errors of the numerical scheme rather than the gradient extraction method. Moreover, the discrete adjoint approach garantees the numerical consistency of the adjoint operator. The continuous adjoint approach however, while garanteeing the consistency at the continuous level, relies on independent integrations of the primal and adjoint problems. As a consequence, inconsistencies may arise which can, as documented in related studies, lead to higher errors in the gradient prediction and in some cases even instability of the results~\cite{Hekmat2016}. The forward sensitivity and continuous adjoint predictions are very close. This is expected, since in both cases, the analytical expression is extracted by applying linear perturbation to the primal problem in continuous form. It should be noted that the finite difference results vary based on the chosen value of $\epsilon$ in $(F(x+\epsilon) - F(x))/\epsilon$. The optimized value of $\epsilon$ was selected here by comparing to the results of the other sensitivity extraction methods. This highlights one of the disadvantages of the finite difference strategy.

\begin{figure}
\centering
{\includegraphics[width=0.5\textwidth]{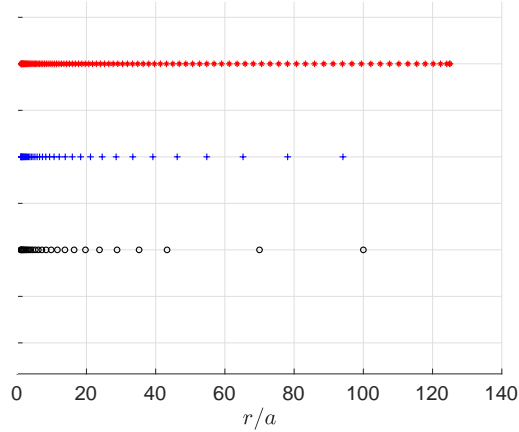}}
\caption{Comparison of the grids used in the different approaches: primal ({\color{black}$\circ$}), continuous adjoint ({\color{red} *}), and forward sensitivity ({\color{blue} +}) problems.}
\label{fig:figure3}
\end{figure}

As previously mentioned, while the grid used in the discrete adjoint is the same as that of the primal problem, the grid of the continuous adjoint solution is different. Figure~\ref{fig:figure3} compares the grids of the different gradient extraction strategies including forward sensitivity. For clarity, only one out of eight grid points is displayed. The total numbers of grid points are $N_{grid}=[769,313,233]$ for continuous, discrete adjoints, and forward sensitivity, respectively. In all cases, the grid points  accumulate at the particle surface, due to the presence of the heterogeneous reactions. While the distance between the grid points increases in both discrete and forward sensitivity methods when moving away from the particle surface, the continuous adjoint shows accumulation of grid points at the far stream boundary, due to the backward integration of the continuous adjoint problem, derived in~\ref{app:adj}. The non-linearity of the problem requires the use of a checkpointing algorithm. In the case of the discrete adjoint, since the grid matches that of the primal problem, the solution is available at each checkpoint. However, due to the differences in the grid distributions in the continuous adjoin approach, an interpolant (cubic splines, in this case) is used in between the checkpoints. As a consequence, the interpolation gives rise to significant errors in the gradient computed using the continuous adjoint approach. This error is not only due to the accuracy of the interpolant, but rather to the large discrepancies between the two grids, as shown in figure~\ref{fig:figure3}, which result in inaccuracies in the construction of the interpolant. 

\section{Results}
\label{sec:results}

In this section, various quantities of interest (QoI's) are identified, locally or spatially distributed, and their sensitivities with respect to the existing combustion parameters are extracted. 
In addition, these sensitivities are extracted considering two different environments, as illustrated in Table \ref{tab:table1}, and are compared. Oxy environment is of interest in coal combustion, as one of viable methods to reduce $\textrm{CO}_2$ emissions~\cite{Chen2012}. In the oxy environment, air is replaced by pure oxygen ($\textrm{O}_2$) or a mixture of oxygen with recycled flue gas, generating high $\textrm{CO}_2$ concentrations, and causing the combustion process to change significantly. The conditions for the two environments in Table~\ref{tab:table1} closely match the conditions imposed in the simulation of Farazi \textit{et al.}~\cite{Farazi2017}.  
\begin{table}[h]
\begin{center}
\begin{tabular}{lccccc}
 Test case & Diluent & $\mathrm{Y}_{\textrm{O}_2}^{\infty}$ & $\mathrm{Y}_{\textrm{CO}}^{\infty}$ & $\mathrm{Y}_{\textrm{CO}_2}^{\infty}$ & $\mathrm{Y}_{\textrm{N}_2}^{\infty}$ \\ 
 Air-atmosphere&  $\textrm{N}_2$ & 0.326 & 0.0 & 0.105 & 0.569 \\  
 Oxy-atmosphere& $\textrm{CO}_2$ & 0.262  & 0.0 & 0.738 & 0.0   
\end{tabular}
\end{center}
\caption{Environments considered for sensitivity analysis. }  
\label{tab:table1}
\end{table}

\begin{figure}[h]
\centering
 \subfloat[Air atmosphere]{\includegraphics[width=0.45\textwidth]{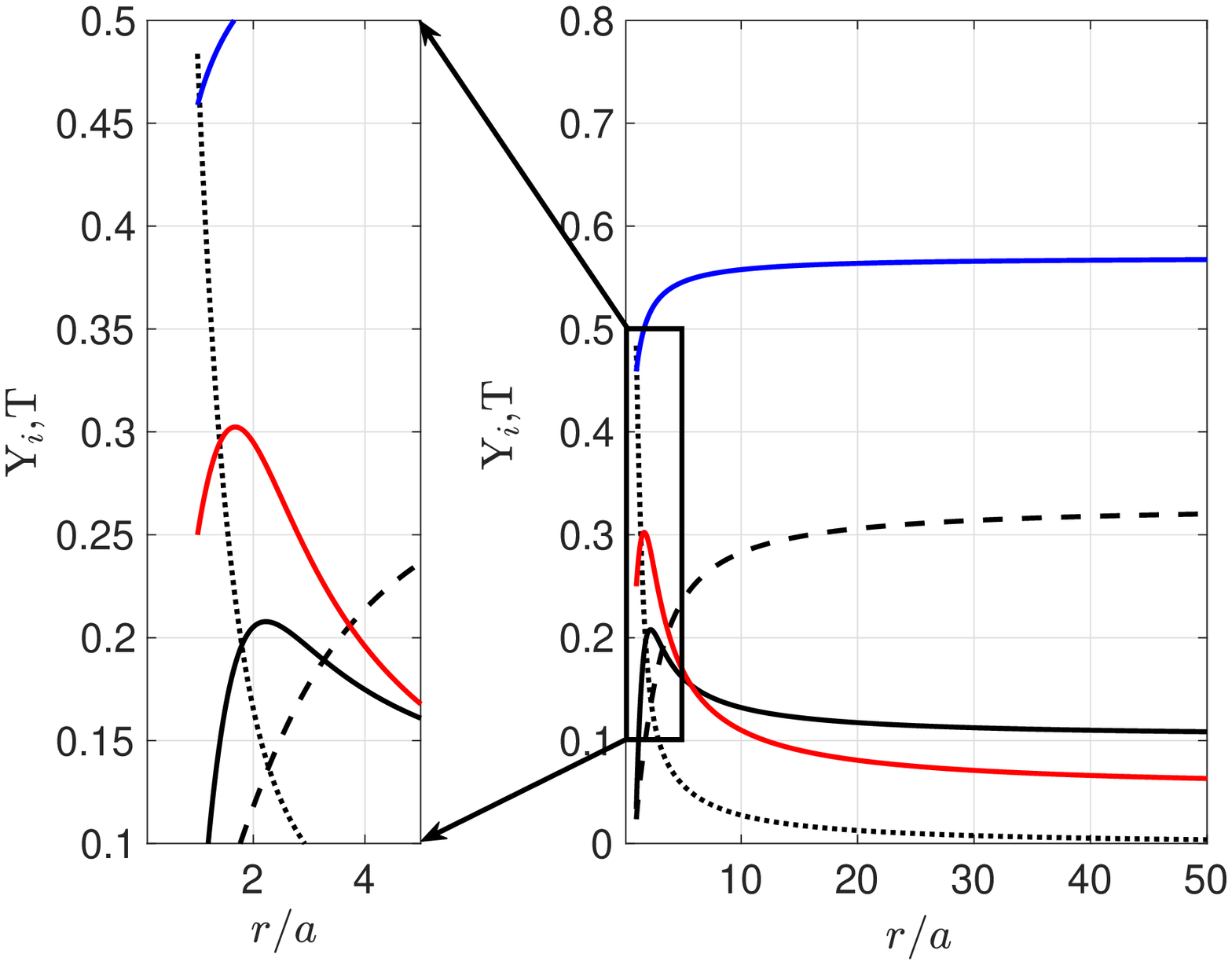}}
  \subfloat[Oxy atmosphere]{\includegraphics[width=0.45\textwidth]{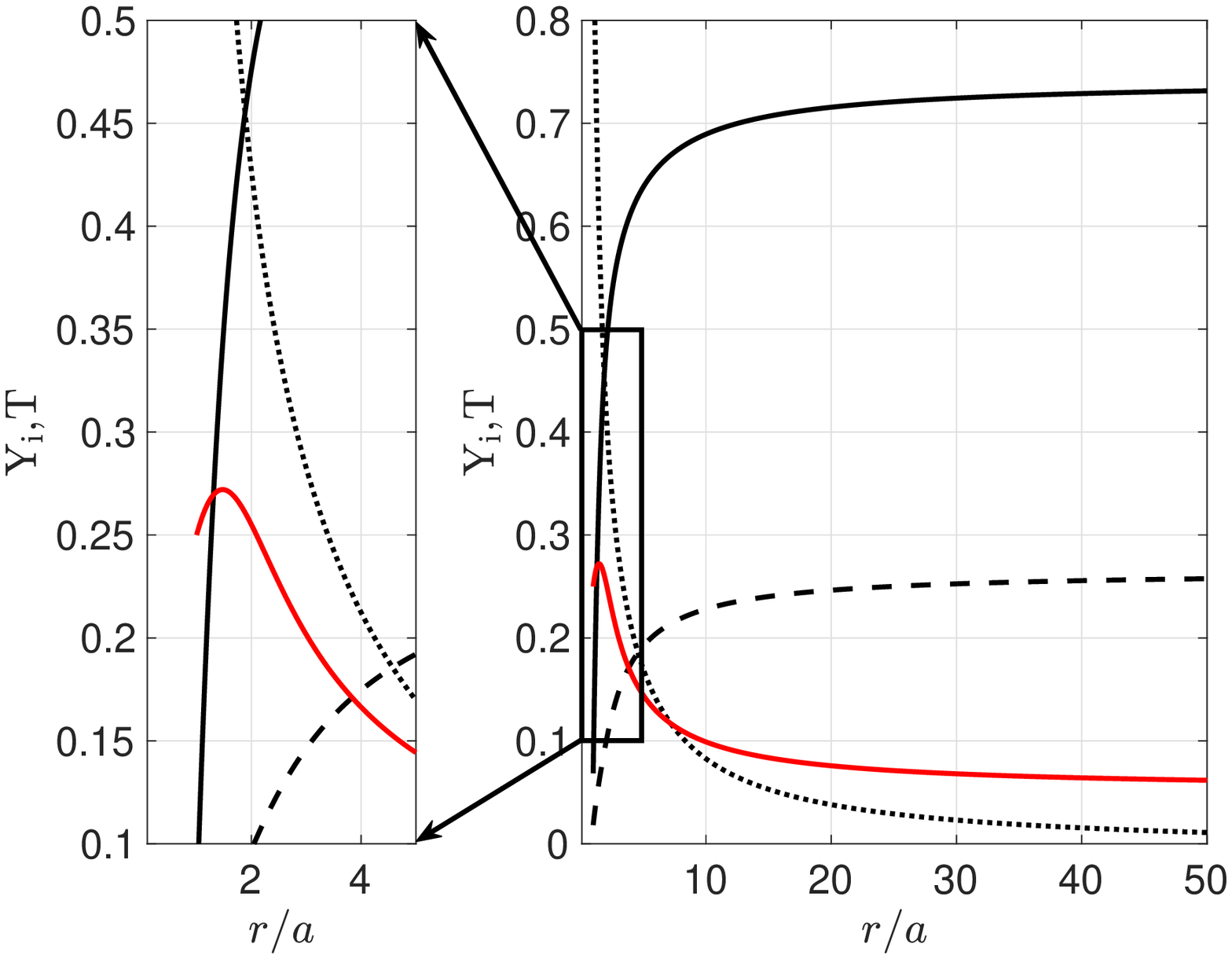}}
\caption{Evolution of mass fractions and temperature in the presence of a flame in both air and oxy atmospheres for $\textrm{Da}^g=2500$, $ \mathrm{Da}^s_1=6$, and $\gamma=3$:  - - -, $\mathrm{Y}_{\textrm{O}_2}$; {\color{blue}---}, $\mathrm{Y}_{\textrm{N}_2}$; {\color{red} ----}, ${\textrm{T}}$; -----, $\mathrm{Y}_{\textrm{CO}_2}$; $\cdots$, $\mathrm{Y}_{\textrm{CO}}$. }
    \label{fig:figure4}
 \end{figure}

While the asymptotic limits $\mathrm{Da}^g = 0$ and $\infty$ provide analytical solutions that can be used to validate the models/numerical methods, in real configurations, the reactions in the gas phase can not be ignored and the results obtained in the frozen limit are not applicable. In what proceeds, the focus is therefore set on finite $\mathrm{Da}^{g}$ values.  One such case for both air and oxy atmospheres is shown in figure~\ref{fig:figure4}, with the following parameter values: $a = 130\; [\mu m]$, $p_0 = 1$ [\textit{atm}], $\mathrm{Y}_{\mathrm{H_2 O}}=0.001$, and the corresponding Damk\"ohler numbers,  $\mathrm{Da}^{g} = 2500$, $\mathrm{Da}^s_1 = 6$, and $\gamma = 3$. In this case, a flame is present in the gas phase and its location is highlighted by the profiles of temperature and carbon dioxide mass fraction reaching their maximal value, at $r/a= 1.675$ for air atmosphere, and $r/a=1.475$ for oxy atmosphere. The location of the maximum temperature, where substantial heat release and oxygen consumption occur, can be qualitatively compared to the two-dimensional direct numerical simulations of Farazi \textit{ et al.}~\cite{Farazi2017}. Although the values obtained for maximum temperature, for air atmosphere $\mathrm{T}_{max} = 2140 \; [K]$, and for oxy-atmpophere $\mathrm{T}_{max} = 1910.5 \; [K]$, compare well to those of the two-dimensional simulations~\cite{Farazi2017}, a shift in the flame position is observed. This shift can be attributed to the presence of a free-stream velocity  in the simulations of Farazi \textit{ et al.}~\cite{Farazi2017}, advecting the gas away from the particle surface. Since the system is diffusion dominant, the behaviour of the diluent species, whether it is a reactive species like carbon dioxide in oxy environment, or an inert species like nitrogen in the air environment, is similar, as illustrated in figures~\ref{fig:figure4}(a) and~\ref{fig:figure4}(b). On the other hand, for the air atmosphere, carbon dioxide behaves similar to the temperature profile, and can therefore be used to identify ignition in the gas phase.

\begin{figure}
\centering
  \subfloat[Air atmosphere]{\includegraphics[width=0.45\textwidth]{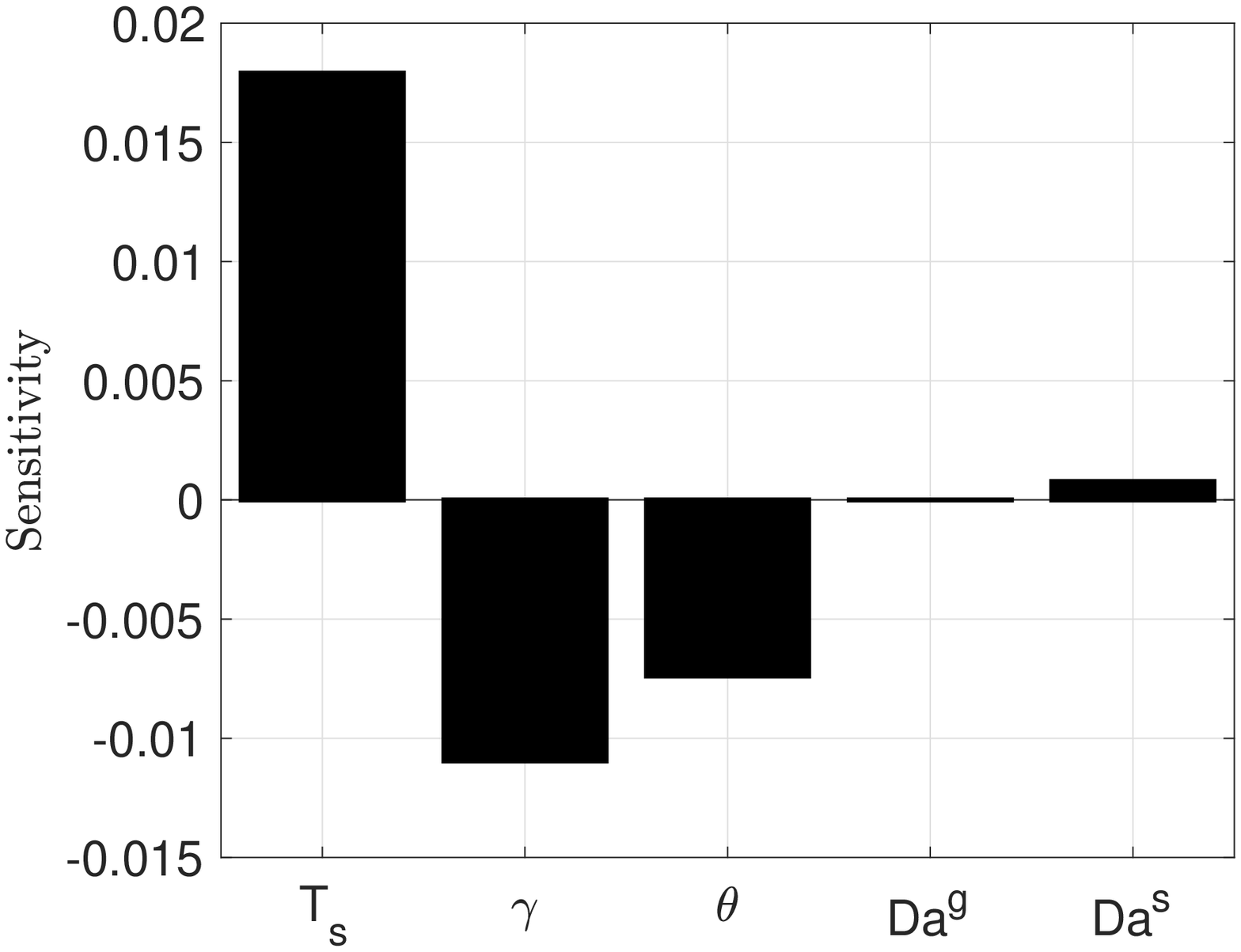}}
  \subfloat[Oxy atmosphere]{\includegraphics[width=0.45\textwidth]{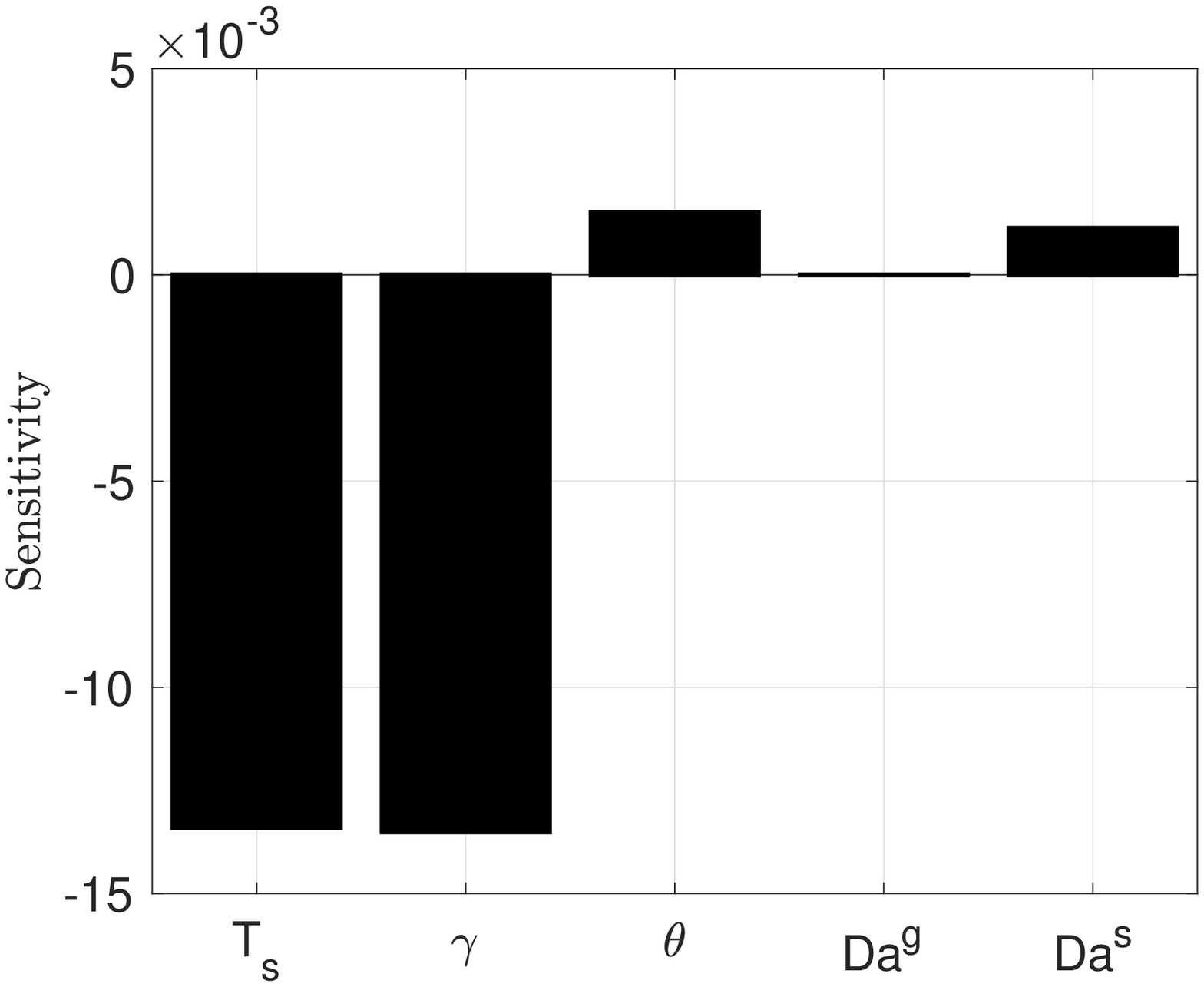}}
\caption{Burning rate sensitivity at operating values, $\mathrm{Da}^{g} = 2500$, $\mathrm{Da}^s_1 = 12$, $\mathrm{{T_s}} = 0.25$ and $\gamma = 0.1$ .}
    \label{fig:figure5}
 \end{figure}

The burning rate, $\dot{M}$, which in this study represents a local quantity, has obvious physical importance, since it determines the value of Stefan flow velocity imposed on the particle surface due to char combustion, and is an indication of the non-linear interaction between flow and chemistry. This value has also mathematical significance, since it corresponds to the eigenvalue of the dynamical system. Therefore, the sensitivities of $\dot{M}$  with respect to various model parameters are extracted for both atmospheres, and the results are shown in figure~\ref{fig:figure5}. In both cases, the burning rate shows the highest sensitivity to the surface parameters: particle temperature and $\gamma$. This result is expected, since $\mathrm{T_s}$ controls the consumption of $\mathrm{O}_2$ and $\mathrm{CO}_2$ at the particle surface, also reported by Makino~\cite{Makino1992}, and helps overcoming the activation energy of the heterogeneous reactions. This dominance persists even when the far-field composition is altered. The gas Damk\"ohler number is the least sensitive parameter, as far as the burning rate is concerned, and this lack of sensitivity can be easily explained by the dependency of the burning rate mainly on the surface reactions, causing the homogeneous reactions to have little impact on the predicted Stefan flow velocity. In addition, the gas Damk\"ohler number is already quite large, leading to the conclusion that gas phase reactions are mixing-controlled. For smaller $\textrm{Da}^g$, this might change, since the homogeneous reaction provides thermal energy to the surface. Note that this result does not indicate that the heterogeneous and homogeneous reactions can be considered independently, the equations are in fact non-linearly coupled. However, it can lead to the conclusion that perturbations on the parameters controlling the intensity of the homogeneous reaction do not cause large variations in the burning rate. As far as the activation energy of the gas phase reaction, $\theta$, is concerned, the sensitivities appear with opposite signs, as the far-field composition is altered. This can be explained by recognising that the selected operating condition for the air atmosphere results in a case with a flame in the gas phase, while for the oxy atmosphere, no flame is present. For cases with $\gamma \ll 1$, such as the two cases here, the char oxidation reaction (defined as reaction~\ref{R1} in section~\ref{sec:GE}) indeed dominates the burning rate. As a result, increasing the activation energy in the no flame region (oxy atmosphere), weakens the ignition possibility in the gas phase, and the gradient of oxygen mass fraction at the surface becomes positive, leading to a positive sign in the burning rate sensitivity. On the other hand, for air atmosphere, augmenting the activation energy of the gas phase in the presence of the flame, leads to an increase in oxygen consumption in the gas phase, causing a negative gradient of oxygen mass fraction at the surface, and consequently, a negative burning rate sensitivity. Similar behaviour can be seen for the sensitivities with respect to surface temperature, since changes in the surface temperature leads to similar behaviour in the gradient of oxygen mass fraction on the surface, positive values for the air and negative values for the oxy atmosphere.

Extracted sensitivities for a spatially-distributed quantity, the total heat release (not shown here), suggest trends similar to the burning rate's. This quantity is of interest since, in most applications, the amount of energy extracted from the system can be easily measured. On the other hand, the total heat release also has mathematical significance, since it represents the transcendental non-linearity of the system. Based on these local results, the gradients with respect to surface parameters will be the focus the following sensitivity analysis.

While local sensitivities, such as those presented in figure~\ref{fig:figure5}, provide valuable information on the influence of certain parameters on the selected quantity of interest, they are dependent on the local conditions of the steady state solution. Experiments and numerical simulations have shown that, various starting conditions can, for example, influence the temperature at the surface of the particle in the steady state limit, and hence alter the local conditions of the system, for which these sensitivities are extracted. Therefore, in order to provide a more comprehensive picture, these sensitivities should be extracted for all possible operating conditions rather than their local counterparts leading to the sensitivity map of the system. Here, the sensitivities are reported for varying surface temperature $\mathrm{T_s}$ and $\gamma$, since the local analysis of figure~\ref{fig:figure5} highlighted these two parameters to be the most sensitive. This parameter space is sampled finely enough that the results remain unchanged with reducing the sampling plane.

\begin{figure}
\centering
  \subfloat[Air atmosphere ]{\includegraphics[width=0.5\textwidth]{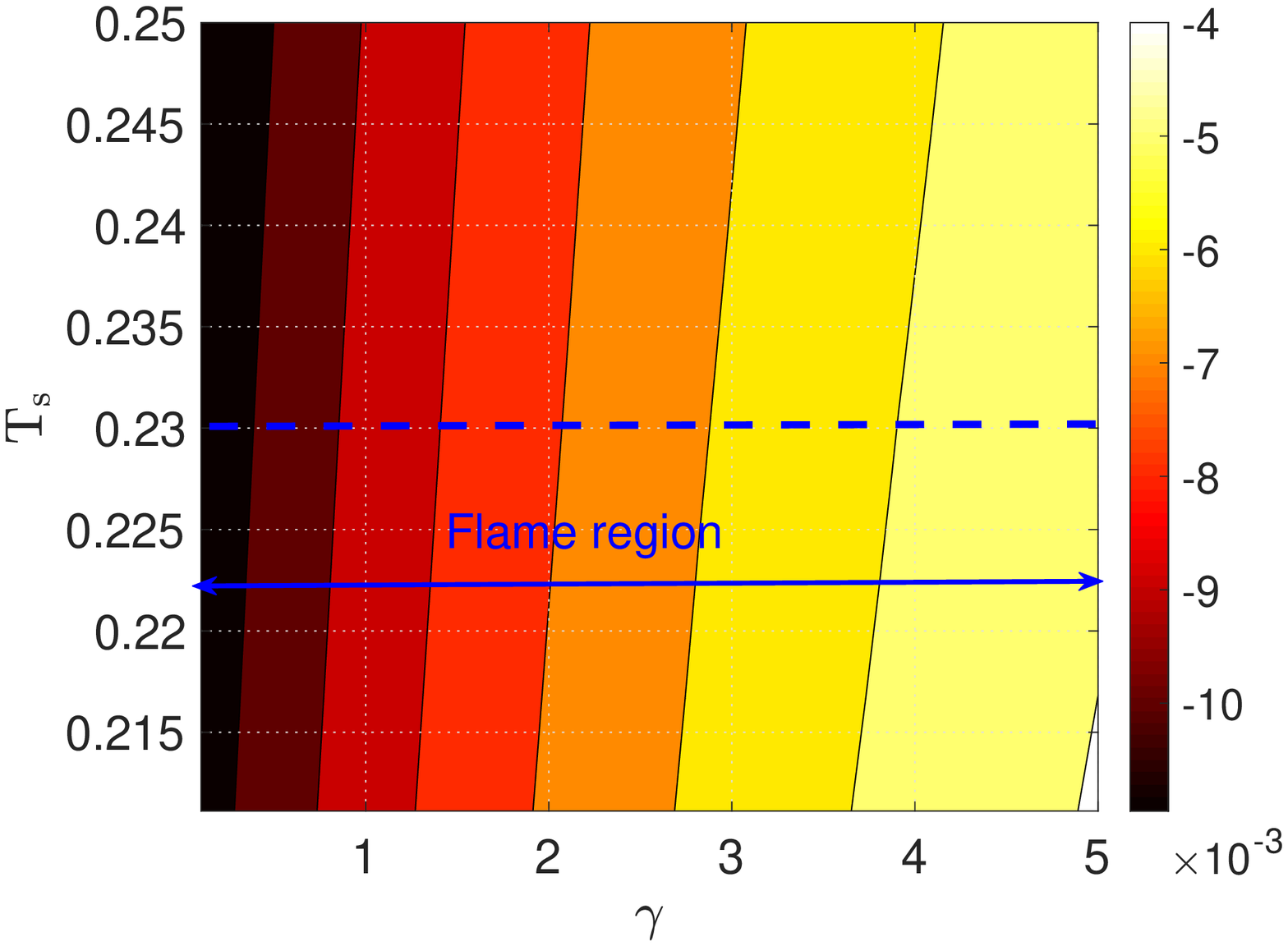}}
  \subfloat[Oxy atmosphere]{\includegraphics[width=0.5\textwidth]{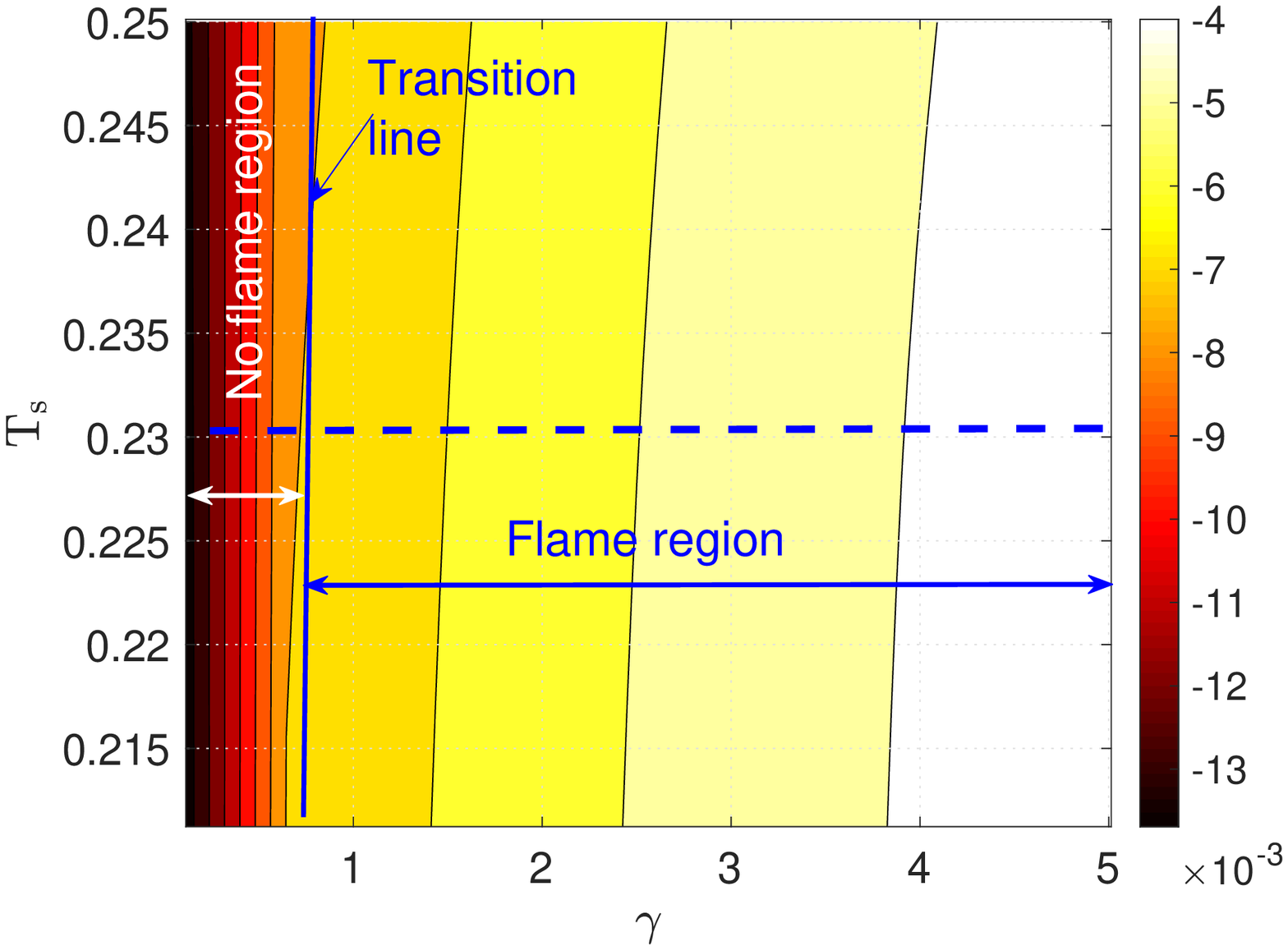}}\\
  \subfloat[Sensitivity reported at $\mathrm{T_s}~=0.23$: -----, air; $\cdots$, oxy.]{\includegraphics[width=0.5\textwidth]{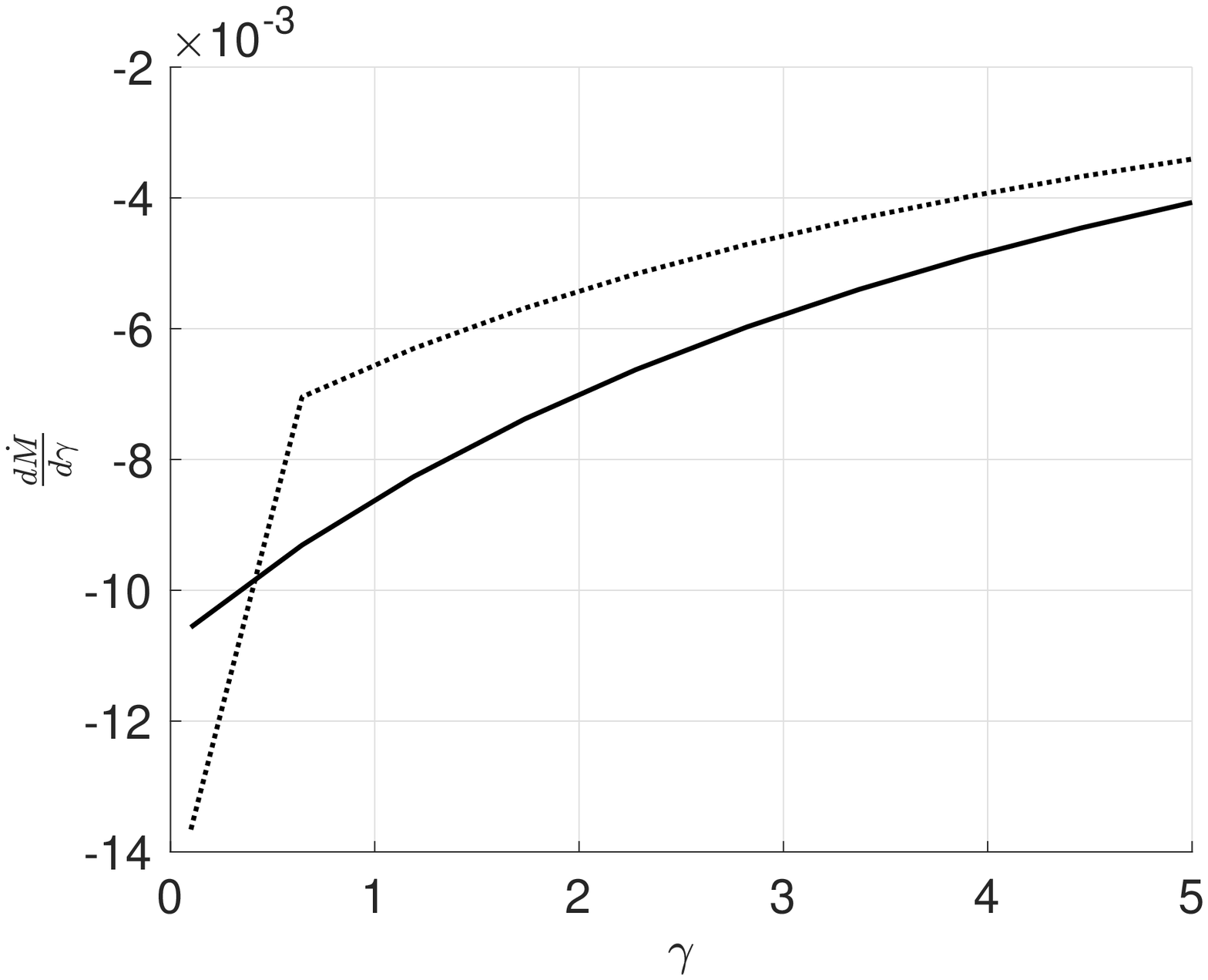}}
\caption{Burning rate sensitivity with respect to $\gamma$, for variable $\mathrm{T_s}$ and $\gamma$, in air and oxy atmospheres.}
    \label{fig:figure6}
 \end{figure}

Figure~\ref{fig:figure6} shows the sensitivity of the burning rate, $\dot{M}$, with respect to $\gamma$ for varying surface temperature and $\gamma$. The range for the surface temperature $\mathrm{T}_s$ is chosen based on the values observed in the experiments of Schiemann \textit{et al.}~\cite{shiemann2014}, whereas, the range for $\gamma$ is selected such that it matches the values investigated by Matalon~\cite{matalon1980,matalon1981}. As illustrated in the figure, the sensitivities of the burning rate with respect to $\gamma$ decrease as the value of $\gamma$ increases. This behaviour seems to be independent of the composition of the surrounding environment, although the level of sensitivities are generally lower (in absolute value) in the oxy atmosphere than in air.  This fact can be explained by analysing the analytical relation of the burning rate gradient with respect to $\gamma$.  In order to extract this expression, first, the equation governing the evolution of burning rate is formally derived, by adding all the source terms of the heterogeneous reactions, leading to the rate of consumption of char by surface reactions, leading to,
 \begin{equation}
    \dot{M}=  \mathrm{Da}^s_1\left (\frac{1-2\alpha}{\alpha}\right)\mathrm{Y}_{\textrm{O}_2}+\mathrm{Da}^s_2\left(1-2\alpha\right)\mathrm{Y}_{\textrm{CO}_2}.
    \label{eq:M}
 \end{equation} 
Taking the derivative of this equation with respect to $\gamma$, gives: 
 \begin{equation}
 \frac{\mathrm{d}\dot{M}}{\mathrm{d}\gamma}= \mathrm{Da}^s_2\left[\left(\frac{1-2\alpha}{\alpha}\right)\left(-\frac{1}{{\gamma}^2} \mathrm{Y}_{\textrm{O}_2}+\frac{1}{\gamma}\frac{\mathrm{d}\mathrm{Y}_{\textrm{O}_2}}{\mathrm{d}\gamma}\right)+(1-2\alpha)\frac{\mathrm{d}\mathrm{Y}_{\textrm{CO}_2}}{\mathrm{d}\gamma}\right].
 \label{eq:dM}
 \end{equation}
This equation highlights nonlinear relation of the burning rate gradient with respect to $\gamma$, which depends on the local values of oxygen mass fraction at the surface of the particle, first term on the right-hand side of the equation, as well as the gradients of oxygen and carbon dioxide mass fractions, second and third terms respectively. The overall sensitivity, reported in figure~\ref{fig:figure6}, therefore, depends on the manner in which each of these terms evolve. Further analysis of each term (not shown here) shows that throughout the chosen parameter range, the local value of oxygen mass fraction has the highest absolute value compared to the rest, and therefore dominates the behaviour of equation~\ref{eq:dM}. As $\gamma$ increases, the local value of oxygen mass fraction at the surface decreases, leading to the decrease of the overall sensitivity. From a physical stand point, since the mass fraction of oxygen on the surface governs the oxidisation of char through reaction~\ref{R1}, it is expected to have a considerable impact on the value of burning rate and its sensitivity. In addition, in the oxy atmosphere, due to the absence of the flame in the low $\gamma$ region, the variation of burning rate with respect to $\gamma$ is more abrupt that in the case with the surrounding air. This behaviour can be better illustrated by comparing the sensitivities, for a variable $\gamma$, at a given surface temperature. The choice of the surface temperature will not impact the analysis, since as shown by figure~\ref{fig:figure6}, this parameter induces a minor variation in the sensitivities, also supported by equation~\ref{eq:dM}, where $\mathrm{T_s}$ does not appear directly. The sensitivities are extracted at $\mathrm{T_s} = 0.23$, marked by dashed lines in figures~\ref{fig:figure6}(a) and~\ref{fig:figure6}(b), for air and oxy atmospheres, and plotted in figure~\ref{fig:figure6}(c). This figure shows that for $\gamma \ll 1$, the sensitivities in the oxy atmosphere are larger in absolute value than in air atmosphere, due to the lack of the flame in the gas-phase in the case of the oxy environment. However, for  $\gamma \gg 1$, where a flame is present in both cases, the oxy atmosphere leads to lower sensitivities overall.  
 \begin{figure}
\centering
  \subfloat[Air atmosphere]{\includegraphics[width=0.5\textwidth]{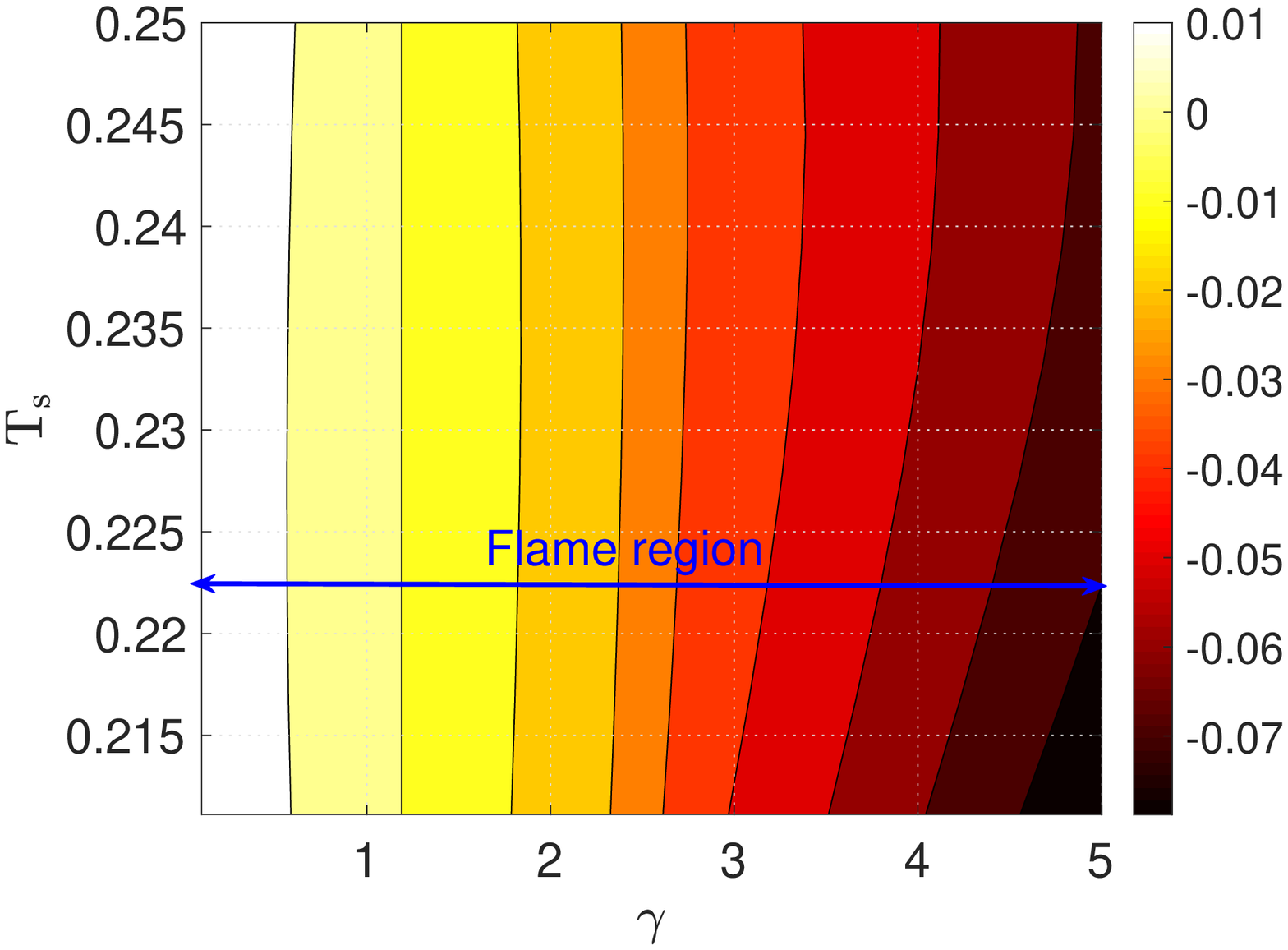}}
  \subfloat[ Oxy atmosphere]{\includegraphics[width=0.5\textwidth]{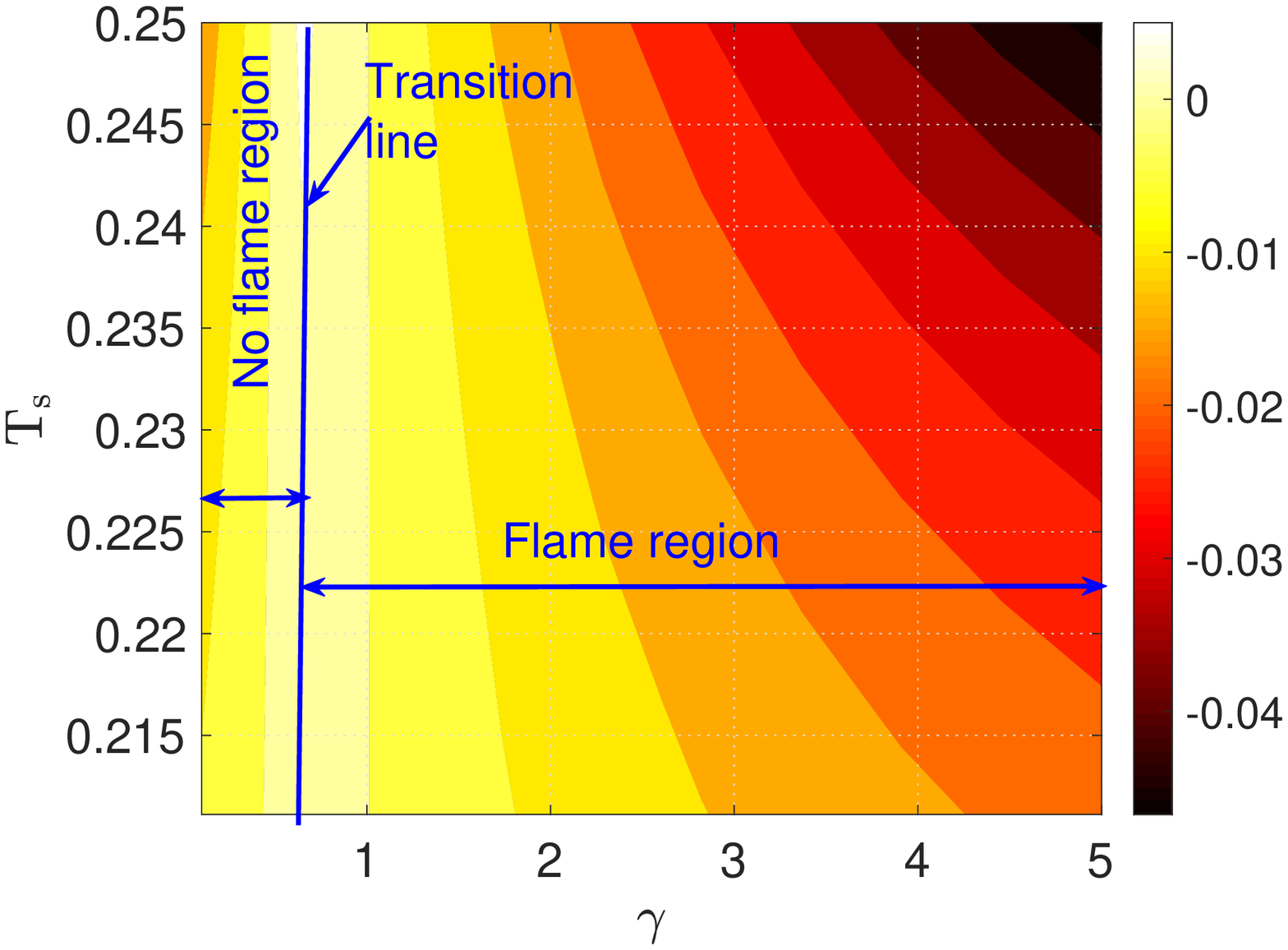}}
\caption{Burning rate sensitivity with respect to surface temperature, $\mathrm{T_s}$, in air and oxy atmospheres. }
\label{fig:figure7}
\end{figure}
Figure~\ref{fig:figure7} shows the sensitivity of the burning rate with respect to the surface temperature, $\mathrm{T_s}$, in the $\mathrm{T_s}$ and $\gamma$ plane. Differentiating equation~\ref{eq:M} with respect to $\mathrm{T_s}$ gives: 
  \begin{equation}
   \frac{\mathrm{d}\dot{M}}{\mathrm{d}\mathrm{T_s}}= \mathrm{Da}^s_2\left[\left(\frac{1-2\alpha}{\alpha}\right)\left(\frac{1}{\gamma}\right)\frac{\mathrm{d}\mathrm{Y}_{\textrm{O}_2}}{\mathrm{d}\mathrm{T_s}}+(1-2\alpha)\frac{\mathrm{d}\mathrm{Y}_{\textrm{CO}_2}}{\mathrm{d}\mathrm{T_s}}\right].
   \label{eq:dM2}
  \end{equation}
As opposed to equation~\ref{eq:dM}, the sensitivity depends not only on the local gradients of mass fractions, but also on the local value of $\gamma$, suggesting that variations of both $\gamma$ and $\mathrm{T_s}$ are expected to affect the extracted sensitivity maps, a fact supported by figure~\ref{fig:figure7}. However, surface temperature seems to have a more dominant effect on the extracted sensitivities of the oxy atmosphere compared to that of the air. This is due to the fact that, the surface temperature in the oxy atmosphere affects the gradient of carbon dioxide mass fraction on the surface more drastically that in air, specifically for higher $\gamma$ values. This figure also shows that, in both air and oxy environments,  the sensitivities increase as $\gamma$ increases. This is in contrast to the behaviour shown in figure~\ref{fig:figure6}. In addition, variations is surface temperatures (figure~\ref{fig:figure7}) seem to lead to higher sensitivities in the burning rate rather than variations in $\gamma$ (figure~\ref{fig:figure6}). While figure~\ref{fig:figure6}(b) shows a monotonic decrease of sensitivities in the oxy environment through the transition from no flame zone to the region containing the flame in the gas phase, figure~\ref{fig:figure7}(b) shows a change of behaviour as the system goes through this transition line. This is mainly due to the absence of the term depending on the local mass-fraction of oxygen mass fraction in equation~\ref{eq:dM2} compared to \ref{eq:dM}. In equation~\ref{eq:dM2}, the only two terms governing the behaviour of the burning rate are the gradient of the oxygen and carbon dioxide mass fractions. For $\gamma \ll 1$, the influence of the first term will dominate. However, as $\gamma$ increases, this influence decreases, until finally at the transition line, the second term becomes larger than the first, leading to the change in the evolution of sensitivities, from no flame to flame regions in the oxy atmosphere. Therefore, in the presence of the flame in the gas phase, both compositions lead to negative sensitivities, whereas for $\gamma \ll 1$, the sensitivities extracted in the air are positive, in contrast to those, negative, in the oxy atmosphere. Similar behaviour was shown in figure~\ref{fig:figure5}, and the same reasoning holds.
 
\begin{figure}
\centering
  \subfloat[Air atmosphere ]{\includegraphics[width=0.5\textwidth]{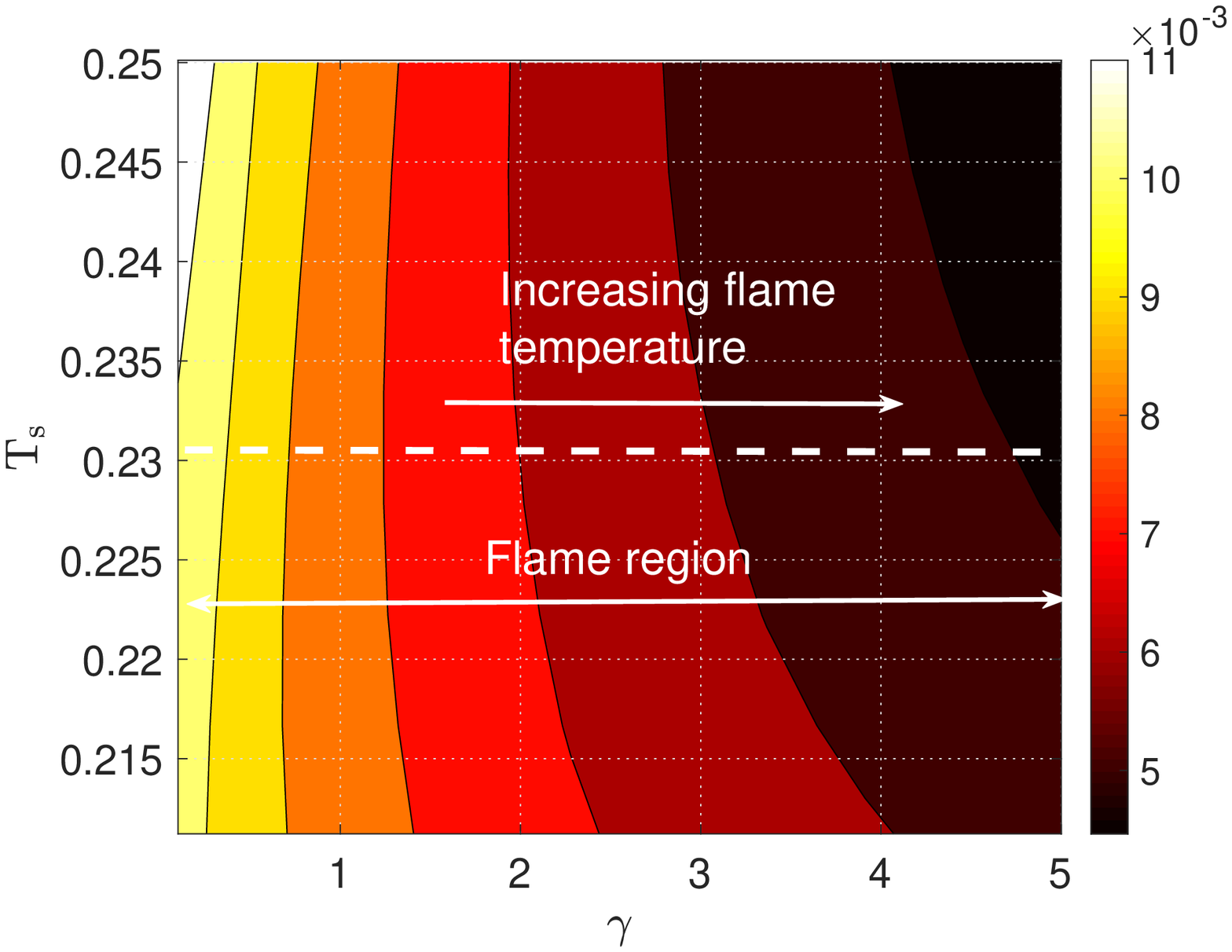}}
  \subfloat[Oxy atmosphere]{\includegraphics[width=0.5\textwidth]{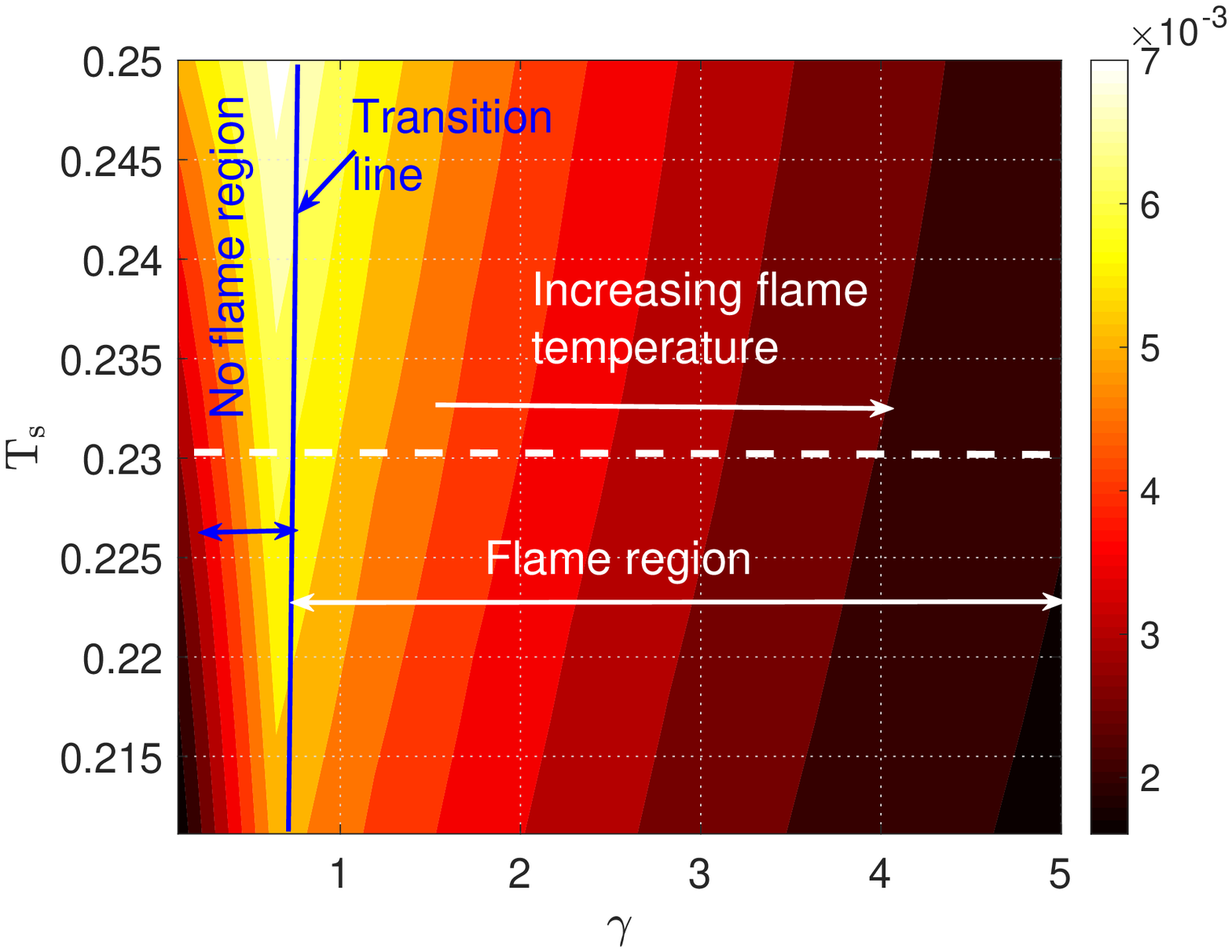}}
\caption{Heat release sensitivity with respect to $\gamma$, for variable $\mathrm{T_s}$ and $\gamma$, in air and oxy atmospheres.}
    \label{fig:figure8}
\end{figure}

While analysing the sensitivities of the burning rate to the surface variables is important, it is of interest to also report the sensitivities of an integrated quantity, such as the heat release~($\int_{a}^{b}\Omega~\mathrm{d
} r$), to the same variables, in order to examine the influence of the choice of QoI on the extracted sensitivities. In this study, the entire computational domain is included in the integral. Figure~\ref{fig:figure8} shows a sensitivity map of the total heat release with respect to $\gamma$, for variable surface temperature and $\gamma$. Similar to figure~\ref{fig:figure6}, as $\gamma$ increases the sensitivities of the total heat release decrease, while the flame temperature~($\mathrm{T_f}$) increases. The comparison between the sensitivities for air and oxy atmospheres shows that the maximum sensitivities are higher for air atmosphere. However, the variation is not monotonous in oxy atmosphere, as opposed to air. While in the region where a flame is present in the gas phase, the sensitivities in both air and oxy environments decrease with the increase of $\gamma$, the behaviour of oxy atmosphere is the opposite as the flame extinguishes. This can be attributed to the fact that in char combustion, due to the presence of reactions on the surfaces, both heterogeneous and homogeneous reactions contribute to the total heat released in the system. Therefore, depending on the presence or absence of a flame in the gas phase, the behaviour of the system can change.  In cases where a flame is present in the gas-phase, due to the higher enthalpy of the homogeneous reaction, and the inverse proportionality between the burning rate value and the flame temperature, the impact of the gas phase reaction on the total heat release dominates. In addition, due to the boundary conditions selected in this problem (surface temperature, $\mathrm{T_s}$, being a known quantity), the contribution of surface reactions to the overall heat-release is predetermined. On the other hand, the heat release in the gas-phase is determined by the overall flame temperature, as a solution to the governing equations. Therefore, it may be expected that the gradient of the heat-release should show dependence on the gradient of the flame temperature. 

 \begin{figure}
\centering
 \subfloat[Air atmosphere ]{\includegraphics[width=0.5\textwidth]{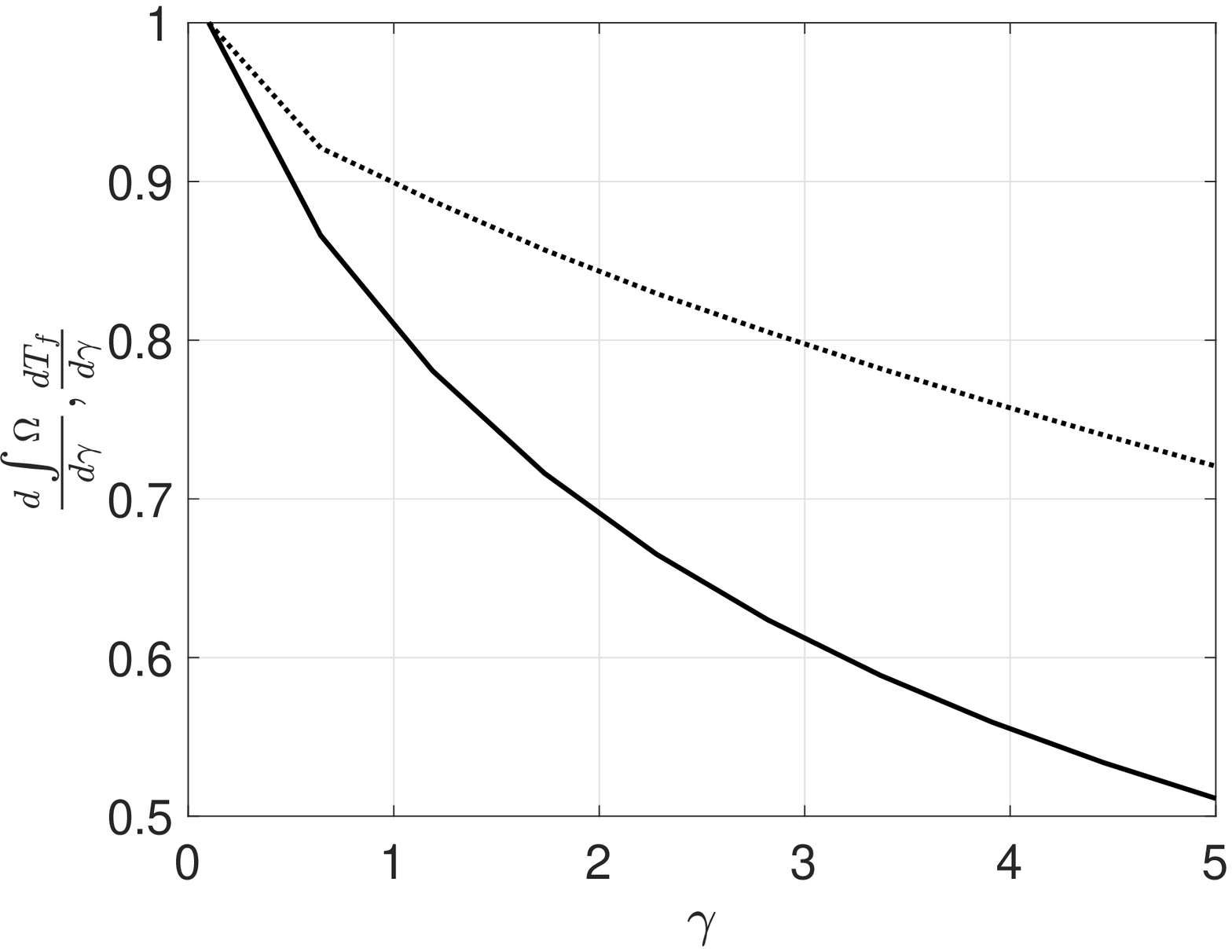}}
  \subfloat[Oxy atmosphere]{\includegraphics[width=0.5\textwidth]{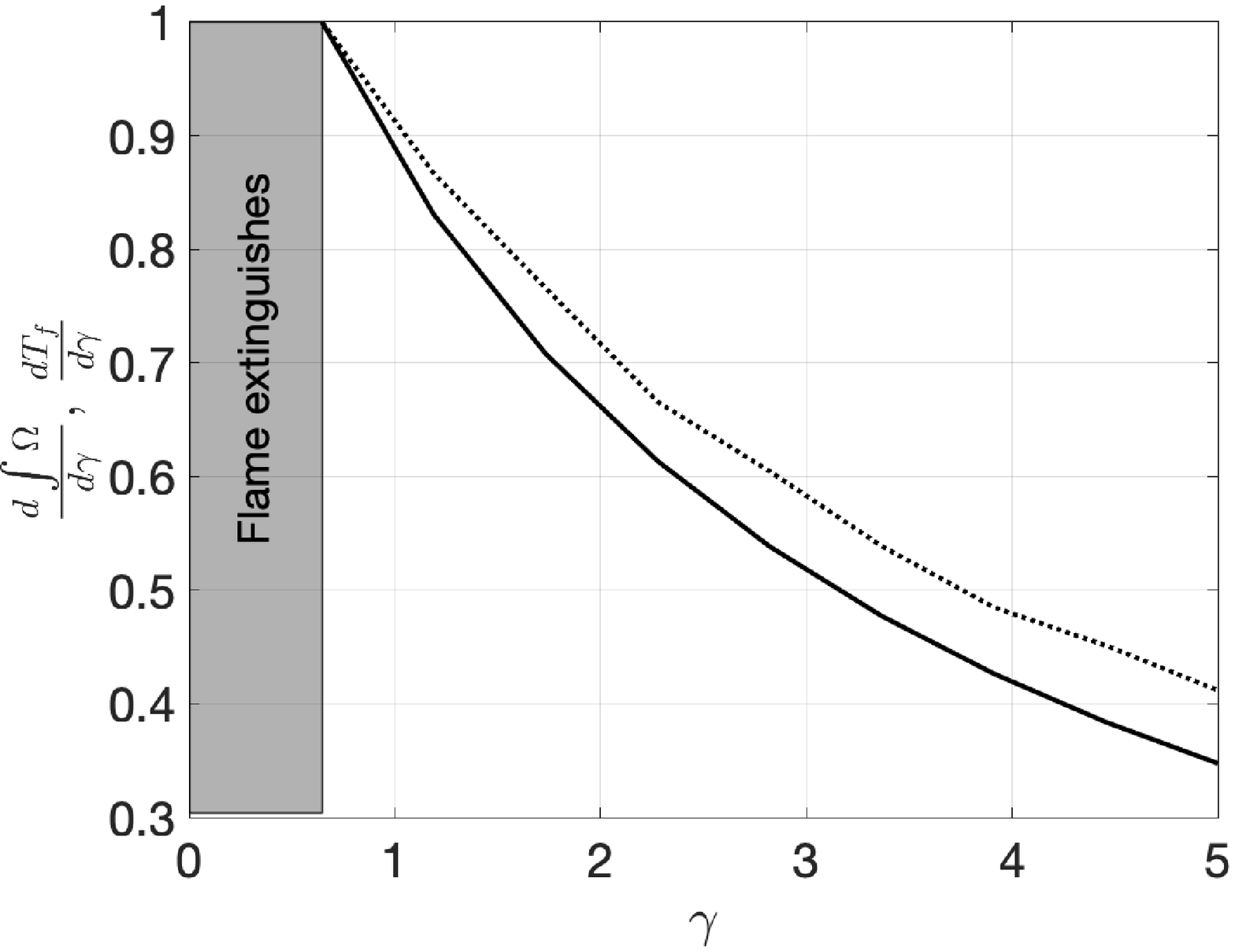}}
\caption{ {\color{blue}}Normalized flame temperature and total heat release sensitivities with respect to $\gamma$ in air and oxy atmospheres at $\mathrm{T_s}=0.23$: ------, total heat release; ........, flame temperature.}
    \label{fig:figure9}
\end{figure}
Figure~\ref{fig:figure9} shows the comparison between the sensitivity of the total heat release in the system and the gradient of the flame temperature for a given surface temperature, $\mathrm{T_s} = 0.23$ (shown by dashed lines in the sensitivity maps of figure~\ref{fig:figure8}). For clarity, both terms have been normalized by the maximum value of each quantity, and plotted for both air and oxy atmospheres. This figure illustrates that both quantities follow the same trend. The maximum gradient appears for the lowest value of $\gamma$ in the air atmosphere. In the oxy atmosphere, however, the values are plotted only for the case where a flame is present in the gas phase. For values lower than the transition value of $\gamma$, the flame extinguishes and moves to the particle surface, where the temperature is set to $\mathrm{T_s}$, and is independent of $\gamma$. In this region, the sensitivity of the total heat release is dominated by the value of oxygen mass fraction at the surface, similar to the sensitivities of the burning rate with respect to $\gamma$, as was explained through equation~\ref{eq:M}. In contrast to the burning rate, the surface temperature seems to affect the extracted sensitivities in both cases, with a higher impact in the oxy atmosphere. Therefore, it will be of interest to also extract the sensitivities with respect to the surface temperature. 

\begin{figure}
\centering
\subfloat[Air atmosphere ]{\includegraphics[width=0.5\textwidth]{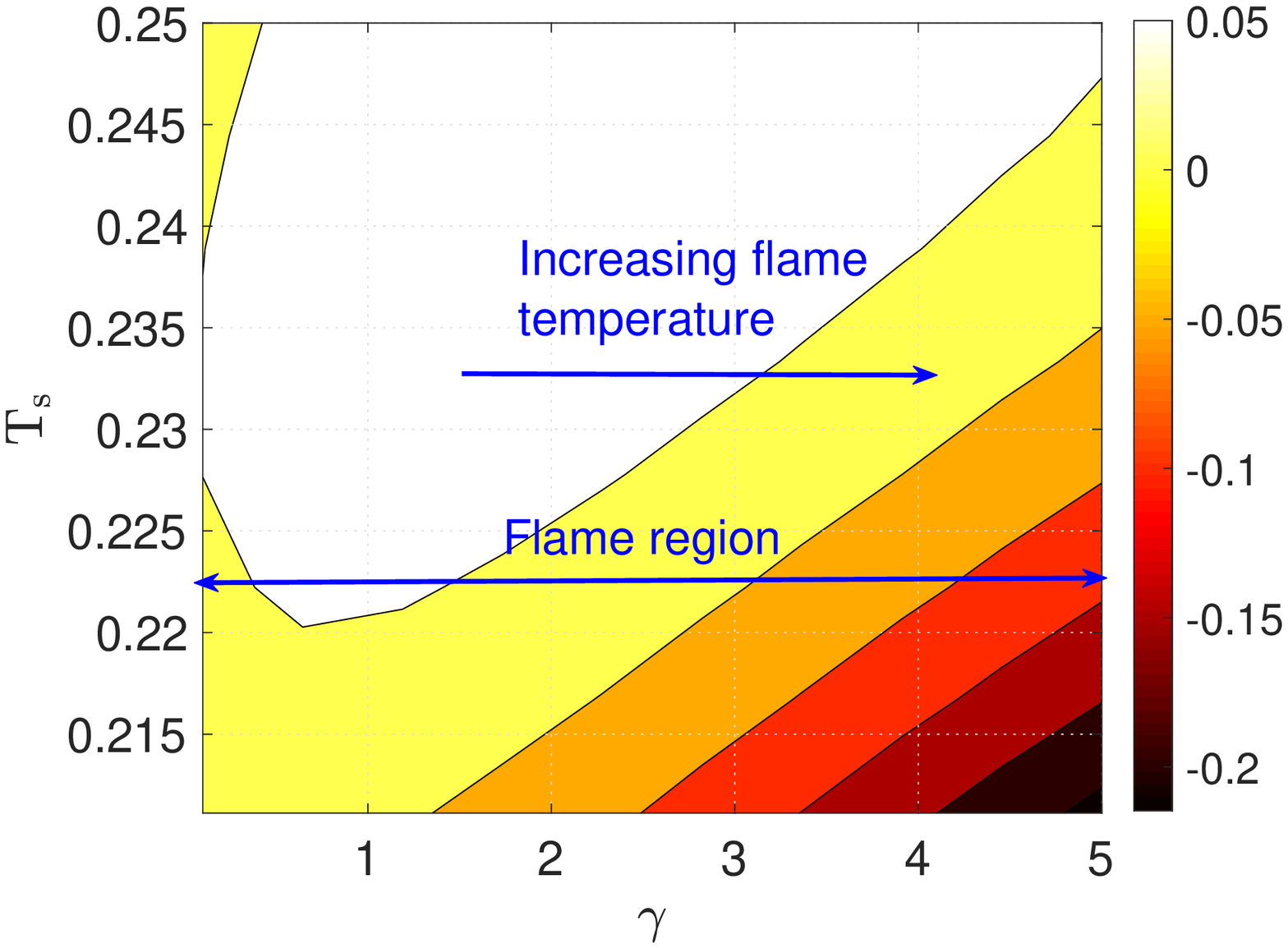}}
\subfloat[Oxy atmosphere]{\includegraphics[width=0.5\textwidth]{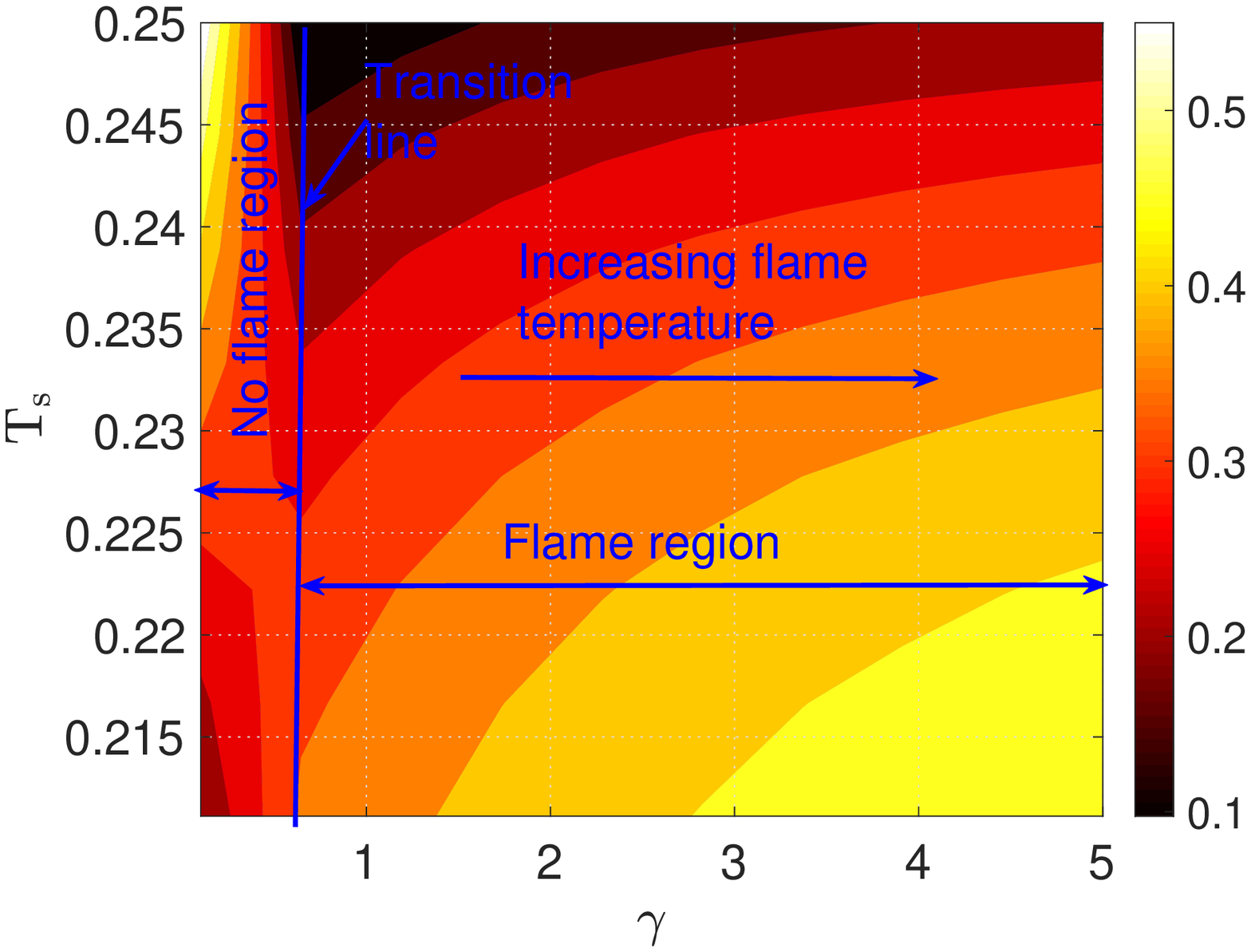}}
\caption{Heat release sensitivity with respect to $\mathrm{T_s}$, for variable $\mathrm{T_s}$ and $\gamma$, in air and oxy atmospheres.}
\label{fig:figure10}
\end{figure}

Figure~\ref{fig:figure10} compares the sensitivities of the total heat releases with respect to surface temperature $\mathrm{T_s}$, in the $\mathrm{T_s}$ and $\gamma$ plane for both air and oxy atmospheres. This figure shows that the maximum sensitivities for oxy atmosphere appears away from the transition line. In the region where a flame is present, the sensitivities increase as $\gamma$ increases and and $\mathrm{T_s}$ decreases, whereas for cases where the flame is at the surface of the particle, the behaviour is the opposite for surface temperatures higher than $\mathrm{T_s} = 0.225$. In the air atmosphere on the other hand, the absolute value of sensitivities generally increase with increasing $\gamma$ and decreasing $\mathrm{T_s}$, following the same trend as the oxy atmosphere, in regions where a flame is present in the gas phase.  The sensitivities of the heat release are much higher with respect to surface temperature than $\gamma$. 


\section{Conclusions and future work}
\label{sec:con}

In this study, we introduce a framework for extracting the steady-state solution and the sensitivities of different QoIs with respect to model parameters of char combustion. Both continuous and discrete adjoint approaches are introduced and compared. A two-step heterogeneous reaction mechanism and a single homogeneous reaction are used to capture the combustion process, as suggested by Matalon~\cite{matalon1980}. The results of the primal problem are then compared to the analytical results of Matalon~\cite{matalon1980} and Makino~\cite{Makino1992} in the frozen limit for verification purposes. Since neither the burning rate nor the composition of the species at the particle surface are known, they are obtained iteratively as the root of the mass balance equations and the saturation conditions. After validating the forward solution, sensitivities extracted using customary gradient extraction techniques (forward sensitivities and finite difference) are compared to those of the continuous and discrete adjoint approaches in the frozen limit. Examining the resulting errors suggests discrete adjoint to be the most accurate formulation for gradient extraction, as far as steady char combustion is concerned. While a frozen limit offers a suitable case for validation purposes, the main interest lies in the configuration where a flame exists in the gas phase. Therefore, the evolution of mass fractions and temperature is simulated in the presence of the flame to account for the gas phase chemistry effect, and sensitivities of various local and spatially distributed QoIs with respect to multiple parameters are extracted and compared, using the discrete adjoint formulation. In addition, due to the importance of the oxy environment in coal combustion, the extracted sensitivities are compared for both air and oxy atmospheres. 

The comparison between the results for a single operating point shows that, regardless of the atmosphere, the system is highly sensitive to the parameters related to the surface reactions, which is expected, since the value of burning rate is directly dependent on the heterogeneous reactions. As the sensitivity of any QoI at a certain operating point does not give an indication about the global behaviour of the gradients, in order to provide a global picture of the sensitivities in the system, sensitivity maps are extracted. Since local analysis shows high sensitivities with respect to the surface parameters, sensitivity maps are also extracted with respect to the parameters governing the surface reactions: surface temperature $\mathrm{T_s}$ and the ratio between the surface Damköhler numbers $\gamma$. For the range of parameters chosen in this study, the oxy environment shows two different characteristics: one dominated by the presence of the flame in the gas phase, and the other seen for low values of $\gamma$, where the flame in the gas-phase extinguishes, and all the dynamics, as well as the flame itself, reside on the surface of the particle, leading to a more dynamic picture in the sensitivity maps. This variation is more apparent in the sensitivities extracted for the total heat release rather than the burning rate. 

The two contributions to the heat release in char combustion are the heterogenous and homogenous reactions. In the presence of the flame, due to the higher enthalpy of the homogenous reaction, the flame temperature dictates the value of heat release and its gradient. However, when the flame moves to the surface of the particle, only the heterogenous reactions govern the value of heat released into the domain. By setting the surface temperature $\mathrm{T_s}$ to a given value, determined by the boundary conditions of the problem, the contribution of the surface reactions are predetermined, and this will influence the extracted sensitivities. Therefore, for a more complete picture, it would be instructive to eliminate the surface temperature from the boundary conditions (resulting in a Robin type boundary condition for the temperature), and allow it to be a solution of the problem. In addition, it should be noted that this study only describes the sensitivities in the stage of char combustion. It would be of interest to study the effect of the devolatilization process on the extracted sensitivities. By adding the volatiles, the kinetics of gas-phase reaction also change, hence changing the overall sensitivities. This effect, however, should be studied by accounting for the unsteadiness of the problem including the ignition process and the transition from devolatilization to char burnout, which will be the subject of future research. 

\section{Acknowledgements}
The authors kindly acknowledge financial support through Deutsch Forschungsgemeinschaft (DFG) through SFB/TRR 129.

\appendix

\section{Derivation of the adjoint equation and gradient formula for two-point boundary value problem}
\label{app:adj}

The notation introduced in section~\ref{sec:Adj} is used below. The derivation follows the original work of Servan \& Petzold~\cite{Serban2003}, with additional details for further clarity.

\subsection{Forward sensitivity of the primal problem}

Let the sensitivity of $\mathbf{u} \left ( r \right )$ with respect to $\mathbf{p}$ be denoted as
$$
\mathbf{u}_\mathbf{p} \left ( r \right ) = \left . \frac{\mathrm{d} \mathbf{u}}{\mathrm{d} \mathbf{p}} \right \vert _ r.
$$
$\mathbf{u}_\mathbf{p}$ is the solution of the linearized problem, referred to as the forward sensitivity problem
$$
\frac{\mathrm{d} \mathbf{u}_\mathbf{p}}{\mathrm{d} r} = \mathbf{F}_\mathbf{u} \mathbf{u}_\mathbf{p} + \mathbf{F}_\mathbf{p},
$$
subjected to the boundary conditions
\begin{equation}
\mathbf{A} \mathbf{u}_\mathbf{p} \left ( a \right ) + \mathbf{B} \mathbf{u}_\mathbf{p} \left ( b \right ) + \mathbf{h}_\mathbf{p} = 0
\label{eq:forwardbc}
\end{equation}
where $\mathbf{A}$ and $\mathbf{B}$ are defined in Eq.~\ref{eqn:bcsensitivity}, $\mathbf{F}_\mathbf{u}$ and $\mathbf{F}_\mathbf{p}$ denote the jacobians of $\mathbf{F}$ with respect to $\mathbf{u}$ and $\mathbf{p}$, and $\mathbf{h}_\mathbf{p}$ the jacobian of $\mathbf{h}$ with respect to $\mathbf{p}$.

\subsection{Adjoint of the primal problem}

This step focuses exclusively on the computation of the gradient of integral-form functionals (Eq.~\ref{eq:integralgrad}), which, is evaluated upon an integration over $\left [ a, s \right ]$. This suggests the use of two adjoint variables : $\boldsymbol{\lambda}_1^s$, the adjoint of the forward sensitivity equation over the interval $\left [ a, s \right ]$, and $\boldsymbol{\lambda}_2^s$ over $\left [ s, b \right ]$. The Lagrange procedure revolves around the addition of the following term to $\mathrm{d} G ^ s / \mathrm{d} \mathbf{p}$ :
$$
\int_a^s \left . \boldsymbol{\lambda}_1^s \right . ^ * \left ( \frac{\mathrm{d} \mathbf{u}_\mathbf{p}}{\mathrm{d} t} - \mathbf{F}_\mathbf{u} \mathbf{u}_\mathbf{p} - \mathbf{F}_\mathbf{p} \right ) \mathrm{d} r + \int_s^b \left . \boldsymbol{\lambda}_2^s \right . ^ * \left ( \frac{\mathrm{d} \mathbf{u}_\mathbf{p}}{\mathrm{d} t} - \mathbf{F}_\mathbf{u} \mathbf{u}_\mathbf{p} - \mathbf{F}_\mathbf{p} \right ) \mathrm{d} r= 0
$$
where the $*$ subscript denotes the conjugate transpose. Integration by parts yields
\begin{equation}
\left [ \left . \boldsymbol{\lambda}_1^s \right . ^ * \mathbf{u}_\mathbf{p} \right ] _ a ^ s + \left [ \left . \boldsymbol{\lambda}_2^s \right . ^ * \mathbf{u}_\mathbf{p} \right ] _ s ^ b - \int_a ^ s \left . \boldsymbol{\lambda}_1^s \right . ^ * \mathbf{F}_\mathbf{p} \mathrm{d} r - \int_s ^ b \left . \boldsymbol{\lambda}_2^ s \right . ^ * \mathbf{F}_\mathbf{p} \mathrm{d} r - \int_a^s \mathbf{u}_\mathbf{p}^* \left ( \frac{\mathrm{d} \boldsymbol{\lambda}_1^s}{\mathrm{d} r} + \mathbf{F}_\mathbf{u} ^ * \boldsymbol{\lambda}_1 ^ s \right ) \mathrm{d} r - \int_s^b \mathbf{u}_\mathbf{p}^* \left ( \frac{\mathrm{d} \boldsymbol{\lambda}_2^s}{\mathrm{d} r} + \mathbf{F}_\mathbf{u} ^ * \boldsymbol{\lambda}_2^s \right ) \mathrm{d} r = 0.
\label{eq:lagrange}
\end{equation}

Upon summation of Eqs.~\ref{eq:integralgrad} and~\ref{eq:lagrange}, it emerges that provided the following three equations hold,
\begin{equation}
\left \{ \begin{aligned}
- \frac{\mathrm{d} \boldsymbol{\lambda}_1^s}{\mathrm{d} r} & = \mathbf{F}_\mathbf{u} ^ * \boldsymbol{\lambda}_1^s + g_\mathbf{u} ^ *, \\
- \frac{\mathrm{d} \boldsymbol{\lambda}_2^s}{\mathrm{d} r} & = \mathbf{F}_\mathbf{u} ^ * \boldsymbol{\lambda}_2^s
\end{aligned} \right .
\label{eq:adjoint}
\end{equation}
and
\begin{equation}
\boldsymbol{\lambda}_1^s \left ( s \right ) - \boldsymbol{\lambda}_2^s \left ( s \right ) = \mathbf{0},
\label{eq:lambdacont}
\end{equation}
the gradient of $G ^ s$ is given by the following formula:
\begin{equation}
\frac{\mathrm{d} G ^ s}{\mathrm{d} \mathbf{p}} = - \left . \boldsymbol{\lambda}_2 ^s \right .^ * \left ( b \right ) \mathbf{u}_\mathbf{p} \left ( b \right ) + \left . \boldsymbol{\lambda}_1 ^s \right .^ * \left ( a \right ) \mathbf{u}_\mathbf{p} \left ( a \right ) + \int_a^s \left ( g _ \mathbf{p} + \left . \boldsymbol{\lambda}_1^s \right .^* \mathbf{F}_\mathbf{p} \right ) \mathrm{d} r + \int_s^b \left . \boldsymbol{\lambda}_2^s \right .^* \mathbf{F}_\mathbf{p} \mathrm{d} r.
\label{eq:integralgrad2}
\end{equation}

If suitable boundary conditions relating $\boldsymbol{\lambda}_1 ^s\left ( a \right )$ and $\boldsymbol{\lambda}_2 ^s\left ( b \right )$ are provided, Eqs.~\ref{eq:adjoint} and~\ref{eq:lambdacont} will form a multi-point boundary value problem for the adjoint variables $\boldsymbol{\lambda}_1^s$ and $\boldsymbol{\lambda}_2^s$. Given the linear dependence of the augmented objective function in the adjoint variables, it is legitimate to assume that the missing boundary conditions too are linear, that is there exist two square matrices $\overline{\mathbf{A}}$ and $\overline{\mathbf{B}}$ such that
\begin{equation}
\overline{\mathbf{A}} \boldsymbol{\lambda}_1^s \left ( a \right ) + \overline{\mathbf{B}} \boldsymbol{\lambda}_2^s \left ( b \right ) = 0
\label{eq:bcadjoint}
\end{equation}
and $\mathbf{P} = \left [ \overline{\mathbf{A}} \vert \overline{\mathbf{B}} \right ]$ has full rank. Note also that given the integral form of $G ^ s$, the conditions were also assumed homogeneous. Eq.~\ref{eq:bcadjoint} states that the boundary conditions $\left ( \boldsymbol{\lambda}_1^s \left ( a \right ), \boldsymbol{\lambda}_2 ^s \left ( b \right ) \right )$ belong to the null space of $\mathbf{P}$.

Since the adjoint methodology aims at avoiding computing forward sensitivities altogether, the aforementioned boundary condition should eliminate $\mathbf{u}_\mathbf{p} \left ( a \right )$ and $\mathbf{u}_\mathrm{p} \left (b \right )$ from Eq.~\ref{eq:integralgrad2} by leveraging the forward sensitivity boundary condition (Eq.~\ref{eq:forwardbc}).
The key to elucidate the missing boundary conditions, and therefore construct $\overline{\mathbf{A}}$ and $\overline{\mathbf{B}}$, is the following theorem : given two $m$ by $2m$ full-rank matrices $\mathbf{P}$ and $\mathbf{Q}$, then
$$
\operatorname{span} \, ( \mathbf{P} ^ * ) = \operatorname{null} \, ( \mathbf{Q} ) \;
\Leftrightarrow \;
\operatorname{span} \, ( \mathbf{Q} ^ * ) = \operatorname{null} \, ( \mathbf{P} ).
$$

In other words, posing $\mathbf{Q} = \left [ - \mathbf{A} \vert \mathbf{B} \right ]$, if $\mathbf{P}$ (and therefore, $\overline{\mathbf{A}}$ and $\overline{\mathbf{B}}$) is constructed such that its row space matches the null space of $\mathbf{Q}$, then since $\left ( \boldsymbol{\lambda}_1^s \left ( a \right ), \boldsymbol{\lambda}_2^s \left ( b \right ) \right )$ belongs to the null space of $\mathbf{P}$, there exists $\boldsymbol{\alpha}$ such that
$$
\left ( \begin{array}{c} \boldsymbol{\lambda}_1^s \left ( a \right ) \\ \boldsymbol{\lambda}_2^s \left ( b \right ) \end{array} \right ) = \left [ \begin{array}{c} - \mathbf{A} ^ * \\ \mathbf{B} ^ * \end{array} \right ] \boldsymbol{\alpha}.
\label{eq:alpha}
$$
As will be shortly seen, if the missing adjoint boundary conditions are be constructed such that
\begin{equation}
\operatorname{span} \, \left ( \left [ \begin{array}{c} \overline{\mathbf{A}} ^ * \\ \overline{\mathbf{B}} ^ * \end{array} \right ] \right ) = \operatorname{null} \, ( \left [ - \mathbf{A} \vert \mathbf{B} \right ] ),
\label{eq:construction}
\end{equation}
then substitution of Eq.~\ref{eq:alpha} in the first two terms of the right-hand side of Eq.~\ref{eq:integralgrad2} yields
$$
\left . \boldsymbol{\lambda}_1^s \right . ^ * \left ( a \right ) \mathbf{u}_\mathbf{p} \left ( a \right ) - \left . \boldsymbol{\lambda}_2^s \right . ^ * \left ( b \right ) \mathbf{u}_\mathbf{p} \left ( b \right ) = - \boldsymbol{\alpha} ^ * \left [ \mathbf{A} \mathbf{u}_\mathbf{p} \left ( a \right ) + \mathbf{B} \mathbf{u}_\mathbf{p} \left ( b \right ) \right ]
$$
where the second factor in the right-hand side matches the first two terms in the sensitivity boundary condition equation~\ref{eq:forwardbc}, which, can therefore be used to eliminate $\mathbf{u}_\mathbf{p} \left ( a \right )$ and $\mathbf{u}_\mathbf{p} \left ( b \right )$. One finally finds
$$
\frac{\mathrm{d} G ^ s}{\mathrm{d} \mathbf{p}} = -\boldsymbol{\alpha} ^ * \mathbf{h}_\mathbf{p} + \int_a^s \left ( g _\mathbf{p} + \left . \boldsymbol{\lambda}_1^s\right . ^* \mathbf{F}_\mathbf{p} \right ) \mathrm{d} r + \int_s^b \left . \boldsymbol{\lambda}_2^s \right .^* \mathbf{F}_\mathbf{p} \mathrm{d} r
$$
where
$$
\boldsymbol{\alpha}  = \left ( \mathbf{A} \mathbf{A} ^ * + \mathbf{B} \mathbf{B} ^ * \right ) ^ {- 1} \left [ \mathbf{B} \boldsymbol{\lambda}_2^s \left ( b \right ) - \mathbf{A} \boldsymbol{\lambda}_1^s \left ( a \right ) \right ].
$$

The matrices $\overline{\mathbf{A}}$ and $\overline{\mathbf{B}}$ derived for the combustion problem of interest are presented in \ref{app:bc}.

\subsection{Forward sensitivity of the adjoint problem}

Let $\boldsymbol{\mu}_i ^ s$ denote the forward sensitivity of $\boldsymbol{\lambda}_i ^ s$ with respect to $s$ ($i = 1, 2$).
$$
\boldsymbol{\mu}_i ^ s = \frac{\mathrm{d} \boldsymbol{\lambda}_i ^ s}{\mathrm{d} s}.
$$
From Eq.~\ref{eq:leibniz}, we find
$$
\left . \frac{\mathrm{d} g}{\mathrm{d} \mathbf{p}} \right \vert _ s = -\boldsymbol{\beta} ^ * \mathbf{h}_\mathbf{p} + g_\mathbf{p} \left ( s \right ) + \int_a^s \left . \boldsymbol{\mu}_1^s \right . ^ * \mathbf{F}_\mathbf{p} \mathrm{d} r + \int_s^b \left . \boldsymbol{\mu}_2^s \right . ^ * \mathbf{F}_\mathbf{p} \mathrm{d} r.
$$
where $\boldsymbol{\lambda}_1 ^ s \left ( s \right ) - \boldsymbol{\lambda}_2 ^ s \left ( s \right )$ was used,
$$
\boldsymbol{\beta} = \left ( \mathbf{A} \mathbf{A} ^ * + \mathbf{B} \mathbf{B} ^ * \right ) ^ {- 1} \left [ \mathbf{B} \boldsymbol{\mu}_2^s \left ( b \right ) - \mathbf{A} \boldsymbol{\mu}_1^s \left ( a \right ) \right ]
$$
and $\left ( \boldsymbol{\mu}_1^s, \boldsymbol{\mu}_2^s \right )$ is the solution of the following multi-point boundary value problem
$$
\left \{ \begin{aligned}
\frac{\mathrm{d} \boldsymbol{\mu}_1^s}{\mathrm{d} r} & = \mathbf{F} _ \mathbf{u} ^ * \boldsymbol{\mu} _ 1 ^ s, \\
\frac{\mathrm{d} \boldsymbol{\mu}_2^s}{\mathrm{d} r} & = \mathbf{F} _ \mathbf{u} ^ * \boldsymbol{\mu} _ 2 ^ s,
\end{aligned} \right .
$$
subject to
$$
\left \{ \begin{aligned}
	\boldsymbol{\mu}_1^s \left ( s \right )  -\boldsymbol{\mu}_2^s \left ( s \right ) - g_\mathbf{u} ^ * \left ( s \right ) & = 0, \\
	\overline{\mathbf{A}} \boldsymbol{\mu}_1^s \left ( a \right ) + \overline{\mathbf{B}} \boldsymbol{\mu}_2^s \left ( b \right ) & = 0.
\end{aligned} \right .
\label{eq:sbc}
$$

The jump condition at $r = s$ in Eq.~\ref{eq:sbc} stems from the differentiation of Eq.~\ref{eq:lambdacont} with respect to $s$, which, results in
$$
\boldsymbol{\mu}_1^s \left ( s \right ) + \left . \frac{\mathrm{d} \boldsymbol{\lambda}_1 ^ s}{\mathrm{d} r} \right \vert_s -\boldsymbol{\mu}_2^s \left ( s \right ) - \left . \frac{\mathrm{d} \boldsymbol{\lambda}_2 ^ s}{\mathrm{d} r} \right \vert_s = 0
$$
that was further simplified using Eq.~\ref{eq:adjoint}.

\subsection{Integration bounds}

Two limit cases are of interest in the present study: $s \to b$ for integral quantities, and $s \to a$ for pointwise quantities. As highlighted below, the multi-point boundary value problems required for adjoint-based the gradient computations reduce to two-point boundary value problems.

\subsubsection{Limit case $s = b$}

In the limit $s \to b$, the adjoint problem reduces to
$$
\left \{ \begin{aligned}
	- \frac{\mathrm{d} \boldsymbol{\lambda} ^ b}{\mathrm{d} r} = \mathbf{F}_\mathbf{u} ^ * \boldsymbol{\lambda} ^ b + g_\mathbf{u}^*, \\
	\overline{\mathbf{A}} ^ * \boldsymbol{\lambda} ^ b \left ( a \right ) +	\overline{\mathbf{B}} ^ * \boldsymbol{\lambda} ^ b \left ( b \right ) = 0,
\end{aligned} \right .
$$
and
$$
\frac{\mathrm{d} G ^ b}{\mathrm{d} \mathbf{p}} = \left ( \mathbf{A} \mathbf{A} ^ * + \mathbf{B} \mathbf{B} ^ * \right ) ^ {- 1} \left ( \mathbf{B} \boldsymbol{\lambda} ^ b \left ( b \right ) - \mathbf{A} \boldsymbol{\lambda} ^ b \left ( a \right ) \right ) \mathbf{h}_\mathbf{p} + \int_a ^ b \left ( g_\mathbf{p} + \left . \boldsymbol{\lambda} ^ b \right . ^ * \mathbf{F}_\mathbf{p} \right ) \mathrm{d} r.
$$

\subsubsection{Limit case $s = a$}

In the limit $s \to a$, the forward sensitivity of the adjoint variable with respect to $s = a$ is the solution of
$$
\left \{ \begin{aligned}
	- \frac{\mathrm{d} \boldsymbol{\mu} ^ a}{\mathrm{d} r} = \mathbf{F} _ \mathbf{u} ^ * \boldsymbol{\mu} ^ a, \\
	\overline{\mathbf{A}} \boldsymbol{\mu} ^ a \left ( a \right ) + \overline{\mathbf{B}} \boldsymbol{\mu} ^ a \left ( b \right ) + \overline{\mathbf{A}} g _ \mathbf{u} ^ * \left ( a \right ) = 0.
\end{aligned} \right .
$$
The gradient of the pointwise quantity $g \left ( a \right )$ is then
$$
\left . \frac{\mathrm{d} g}{\mathrm{d} \mathbf{p}} \right \vert _ a = \left ( \mathbf{A} \mathbf{A} ^ * + \mathbf{B} \mathbf{B} ^ * \right ) ^ {- 1} \left ( \mathbf{B} \boldsymbol{\mu} ^ a \left ( b \right ) - \mathbf{A} \boldsymbol{\mu} ^ a \left ( a \right ) - \mathbf{A} g _ \mathbf{u} ^ * \left ( a \right ) \right ) + g _ \mathbf{p} \left ( a \right ) + \int_a ^ b \left . \boldsymbol{\mu} ^ a \right . ^ * \mathbf{F} _\mathbf{p} \mathrm{d} r.
$$

\section{Linearized boundary conditions}
\label{app:bc}

\subsection{Forward problem}

The jacobian matrices $\mathbf{A}$ and $\mathbf{B}$ of the boundary conditions, listed in the following order, are provided below. First, at the particle surface, species conservation (4 equations), temperature and saturation condition. Second, in the far field, composition (4 equations) and temperature.

\begin{equation}
\mathbf{A} = \left [ \begin{array}{@{}*{11}{c}@{}}
\dot{M} & \left ( \alpha - 1 \right ) \mathrm{Da}_1 / \alpha & 2 \left ( \alpha - 1 \right ) \mathrm{Da}_2 & 0 & 0 & \mathrm{Y}_{CO} & -1 & 0 & 0 & 0 & 0 \\
0 & \dot{M} + \mathrm{Da}_1 & 0 & 0 & 0 & \mathrm{Y}_{O_2} & 0 & -1 & 0 & 0 & 0 \\
0 & 0 & \dot{M} + \mathrm{Da}_2 & 0 & 0 & \mathrm{Y}_{CO_2} & 0 & 0 & -1 & 0 & 0 \\
0 & 0 & 0 & \dot{M} & 0 &\mathrm{Y}_{N_2} & 0 & 0 & 0 & -1 & 0 \\
0 & 0 & 0 & 0 & 1 & 0 & 0 & 0 & 0 & 0 & 0 \\
1 & 1 & 1 & 1 & 0 & 0 & 0 & 0 & 0 & 0 & 0 \\
0 & 0 & 0 & 0 & 0 & 0 & 0 & 0 & 0 & 0 & 0 \\
0 & 0 & 0 & 0 & 0 & 0 & 0 & 0 & 0 & 0 & 0 \\
0 & 0 & 0 & 0 & 0 & 0 & 0 & 0 & 0 & 0 & 0 \\
0 & 0 & 0 & 0 & 0 & 0 & 0 & 0 & 0 & 0 & 0 \\
0 & 0 & 0 & 0 & 0 & 0 & 0 & 0 & 0 & 0 & 0
\end{array} \right ]
\end{equation}

\begin{equation}
\mathbf{B} = \left [ \begin{array}{@{}*{11}{c}@{}}
0 & 0 & 0 & 0 & 0 & 0 & 0 & 0 & 0 & 0 & 0 \\
0 & 0 & 0 & 0 & 0 & 0 & 0 & 0 & 0 & 0 & 0 \\
0 & 0 & 0 & 0 & 0 & 0 & 0 & 0 & 0 & 0 & 0 \\
0 & 0 & 0 & 0 & 0 & 0 & 0 & 0 & 0 & 0 & 0 \\
0 & 0 & 0 & 0 & 0 & 0 & 0 & 0 & 0 & 0 & 0 \\
0 & 0 & 0 & 0 & 0 & 0 & 0 & 0 & 0 & 0 & 0 \\
1 & 0 & 0 & 0 & 0 & 0 & 0 & 0 & 0 & 0 & 0 \\
0 & 1 & 0 & 0 & 0 & 0 & 0 & 0 & 0 & 0 & 0 \\
0 & 0 & 1 & 0 & 0 & 0 & 0 & 0 & 0 & 0 & 0 \\
0 & 0 & 0 & 1 & 0 & 0 & 0 & 0 & 0 & 0 & 0 \\
0 & 0 & 0 & 0 & 1 & 0 & 0 & 0 & 0 & 0 & 0
\end{array} \right ]
\end{equation}

\subsection{Adjoint problem}


In order to express the boundary conditions of the adjoint problem, we define
$$
Z = \mathrm{Y}_{\mathrm{CO}} \left ( a \right ) + \frac{\alpha \dot{M} + \left ( 1 - \alpha \right ) \mathrm{Da} ^ s _ 1}{\alpha \dot{M} + \alpha \mathrm{Da} ^ s _ 1} \mathrm{Y}_{\mathrm{O}_2} \left ( a \right ) + \frac{\dot{M} + 2 \left ( 1 - \alpha \right ) \mathrm{Da} ^ s _ 2}{\dot{M} + \mathrm{Da} ^ s _ 2} \mathrm{Y}_{\mathrm{CO}_2} \left ( a \right ) + \mathrm{Y}_{\mathrm{N}_2} \left ( a \right ).
$$

The matrices $\overline{\mathbf{A}}$ and $\overline{\mathbf{B}}$ are constructed so as to satisfy Eq.~\ref{eq:construction}, that is a basis for the null space of $\mathbf{Q}$ is constructed and used to assemble $\overline{\mathbf{A}}$ and $\overline{\mathbf{B}}$.
The choice of the basis is of course not unique, and the following matrices are but one example :
\begin{equation}
\overline{\mathbf{A}} = \left [ \begin{array}{@{}*{11}{c}@{}}
	\frac{\mathrm{Y}_{\mathrm{O}_2}}{\dot{M} + \mathrm{Da} ^ s _ 1} + \frac{\mathrm{Y}_{\mathrm{CO}_2}}{\dot{M} + \mathrm{Da} ^ s _ 2} + \frac{\mathrm{Y}_{\mathrm{N}_2}}{\dot{M}} & -\frac{\mathrm{Y}_{\mathrm{O}_2}}{\dot{M} + \mathrm{Da} ^ s _ 1} & -\frac{\mathrm{Y}_{\mathrm{CO}_2}}{\dot{M} + \mathrm{Da} ^ s _ 2} & -\frac{\mathrm{Y}_{\mathrm{N}_2}}{\dot{M}} & 0 & 1 & Z & 0 & 0 & 0 & 0 \\
	 \frac{\mathrm{Y}_{\mathrm{O}_2} - Z}{\dot{M} + \mathrm{Da} ^ s _ 1} + \frac{\mathrm{Y}_{\mathrm{CO}_2}}{M + \mathrm{Da}_2 ^ 2} + \frac{\mathrm{Y}_{\mathrm{N}_2}}{\dot{M}} & \frac{Z - \mathrm{Y}_{\mathrm{O}_2}}{\dot{M} + \mathrm{Da} ^ s _ 1} & - \frac{\mathrm{Y}_{\mathrm{CO}_2}}{\dot{M} + \mathrm{Da}_2 ^ 2} & -\frac{\mathrm{Y}_{\mathrm{N}_2}}{\dot{M}} & 0 & 1 & 0 & Z & 0 & 0 & 0 \\
	\frac{\mathrm{Y}_{\mathrm{O}_2}}{\dot{M} + \mathrm{Da} ^ s _ 1} + \frac{\mathrm{Y}_{\mathrm{CO}_2} - Z}{M + \mathrm{Da} ^ s _ 2} + \frac{\mathrm{Y}_{\mathrm{N}_2}}{\dot{M}} & -\frac{\mathrm{Y}_{\mathrm{O}_2}}{\dot{M} + \mathrm{Da} ^ s _ 1} & \frac{Z - \mathrm{Y}_{\mathrm{CO}_2}}{M + \mathrm{Da} ^ s _ 2} & -\frac{\mathrm{Y}_{\mathrm{N}_2}}{M} & 0 & 1 & 0 & 0 & Z & 0 & 0 \\
	\frac{\mathrm{Y}_{\mathrm{O}_2}}{\dot{M} + \mathrm{Da} ^ s _ 1} + \frac{\mathrm{Y}_{\mathrm{CO}_2}}{\dot{M} + \mathrm{Da} ^ s _ 2} + \frac{\mathrm{Y}_{\mathrm{N}_2} - Z}{M} & -\frac{\mathrm{Y}_{\mathrm{O}_2}}{\dot{M} + \mathrm{Da} ^ s _ 1} & -\frac{\mathrm{Y}_{\mathrm{CO}_2}}{\dot{M} + \mathrm{Da} ^ s _ 2} & \frac{Z - \mathrm{Y}_{\mathrm{N}_2}}{\dot{M}} & 0 & 1 & 0 & 0 & 0 & Z & 0 \\
	0 & 0 & 0 & 0 & 0 & 0 & 0 & 0 & 0 & 0 & 1 \\
	0 & 0 & 0 & 0 & 0 & 0 & 0 & 0 & 0 & 0 & 0 \\
	0 & 0 & 0 & 0 & 0 & 0 & 0 & 0 & 0 & 0 & 0 \\
	0 & 0 & 0 & 0 & 0 & 0 & 0 & 0 & 0 & 0 & 0 \\
	0 & 0 & 0 & 0 & 0 & 0 & 0 & 0 & 0 & 0 & 0 \\
	0 & 0 & 0 & 0 & 0 & 0 & 0 & 0 & 0 & 0 & 0 \\
	0 & 0 & 0 & 0 & 0 & 0 & 0 & 0 & 0 & 0 & 0
\end{array} \right ]
\label{eq:Abar}
\end{equation}
and
\begin{equation}
\overline{\mathbf{B}} = \left [ \begin{array}{@{}*{11}{c}@{}}
	0 & 0 & 0 & 0 & 0 & 0 & 0 & 0 & 0 & 0 & 0 \\
	0 & 0 & 0 & 0 & 0 & 0 & 0 & 0 & 0 & 0 & 0 \\
	0 & 0 & 0 & 0 & 0 & 0 & 0 & 0 & 0 & 0 & 0 \\
	0 & 0 & 0 & 0 & 0 & 0 & 0 & 0 & 0 & 0 & 0 \\
	0 & 0 & 0 & 0 & 0 & 0 & 0 & 0 & 0 & 0 & 0 \\
	0 & 0 & 0 & 0 & 0 & 1 & 0 & 0 & 0 & 0 & 0 \\
	0 & 0 & 0 & 0 & 0 & 0 & 1 & 0 & 0 & 0 & 0 \\
	0 & 0 & 0 & 0 & 0 & 0 & 0 & 1 & 0 & 0 & 0 \\
	0 & 0 & 0 & 0 & 0 & 0 & 0 & 0 & 1 & 0 & 0 \\
	0 & 0 & 0 & 0 & 0 & 0 & 0 & 0 & 0 & 1 & 0 \\
	0 & 0 & 0 & 0 & 0 & 0 & 0 & 0 & 0 & 0 & 1
\end{array} \right ].
\label{eq:Bbar}
\end{equation}

With this choice, the adjoint problem is therefore subjected to homogeneous boundary conditions of the following types :
\begin{itemize}
\item At $r = a$, four Robin conditions (one per adjoint species mass fraction) and one Neumann condition (for the adjoint temperature),
\item At $r = b$, one Dirichlet condition (for the adjoint burning rate) and five Neumann conditions (one per adjoint species mass fraction, plus one for the adjoint temperature).
\end{itemize}

\end{document}